\documentclass{elsart}

\usepackage{natbib}

 \usepackage{graphicx}

\usepackage{amssymb}
\newcommand{\fr}{\frac}

\begin{document}

\begin{frontmatter}



\title{On Mechanics and Thermodynamics of a stellar galaxy in a two-component virial system 
and the Fundamental Plane.}


\author{L. Secco}
\address{Department of Astronomy, University of Padova, Padova, Italy}
\ead{secco@pd.astro.it}

\begin{abstract}
The paper confirms the existence of a special configuration 
(among the infinitive number of {\it a priori}
possible virial states) which a $B$ stellar (Baryonic) component
may assume inside a given $D$ dark halo potential well. This satisfies the 
d'Alembert Principle of {\it virtual works} and its typical dimension works as
a scale length (we call {\it tidal radius}) induced on the
gravitational field of the bright
component by the dark one. Its dynamic and thermodynamic properties are
here analyzed  
in connection with the physical reason for the existence of the 
Fundamental Plane for ellipticals and, in general, for two-component virialized systems.
The analysis is performed 
by using two-component models with two power-law density profiles
and two homogeneous cores. The outputs of this kind of models, at the special configuration,
are summarized
and compared with some observable scaling relations for pressure supported
ellipticals. The problem of extending the results to a general class of models
with Zhao (1996) profiles, which are more suitable for an elliptical galaxy system,
is also taken into account.
The virial equilibrium stages of the two-component system have
to occur after a previous violent relaxation phase. If the stellar
$B$ component is allowed to cool slowly its virial evolution consists    
of a sequence of contractions
with enough time to rearrange the virial equilibrium after any step.
The thermodynamic process during the dynamical evolution is so divided into
a sequence of transformations which are irreversible but 
occur between two quasi-equilibrium stages.
Then, it is possible to assign: a mean temperature 
to the whole $B$ component during this quasi-static sequence 
and the entropy variation between two consecutive virial steps.
The analysis allows the conclusion that the induced scale length is a real confinement for the
stellar system. This follows from the application of the
$I^o$ Thermodynamics Principle
under the virial equilibrium constraint, by checking how larger configurations  
turn out to be forbidden, according to the
$II^o$ Thermodynamics Principle.
The presence of this specific border on the space of the baryonic luminous
component has to be regarded as
the physical reason why a stellar galaxy belongs to the Fundamental Plane
(FP)
and why astrophysical objects, with a completely different history 
and formation, but characterized by a tidal radius (as the globular clusters are) 
lie on the same FP.
An other problem addressed is
how this special configuration may be reached. This 
is strictly connected with
the problem of the end state of the collisionless stellar system after
a violent relaxation phase. Even if degeneracy towards the initial
conditions is present on the FP, the mechanic and thermodynamic 
properties of
the special configuration suggest this state may be the best
candidate for the beginning of the $B$ component
virial evolution, and also give a possible explanation for why an elliptical is not
completely relaxed in respect to its dark halo.

\end{abstract}


\end{frontmatter}

\section{Introduction}
As Ogorodnikov (1965) has highlighted, in order to find the most probable phase distribution function
for a stellar system in a stationary state, the phase volume has to be truncated  in both
coordinate and velocity space. While in the velocity space the truncation 
arises spontaneously due to the existence of the velocity of escape, the introduction of a cut-off in the 
coordinate space appears, on one side,
necessary in order to obtain a finite mass $M$ and radius $R$, but, on the
other, very problematic. 

A similar 
difficulty also 
appears on the
thermodynamical side, for which an extensive literature exists ( from: Lynden-Bell
\& Wood, 1968; Horowitz \& Katz, 1978; White \& Narayan, 1987, until, e.g., 
Bertin \& Trenti, 2003, and references therein). By using 
the standard Boltzmann-Gibbs entropy:
\begin{equation}
\label{entr}
{\bf S}=-\int {flnf d^3x d^3v}
\end{equation}
defined by
the {\it distribution function} in the $6-dimensional$ phase space , $f(\vec{x},\vec{v})$
(hereafter $DF$), and looking for
what maximizes the entropy of the same stellar system, the conclusion is: the $DF$
which plays this role in (\ref{entr}) is that of the isothermal sphere. But, the 
maximization of ${\bf S}$, subject to fixed mass $M$ and energy $E$, leads again to a $DF$ that is incompatible with
finite $M$ and $E$ ( see, e.g., Binney
\& Tremaine, 1987, Chapter 4; Merritt 1999, Lima Neto et al. 1999, Marquez et al. 2001, 
and references therein).

It is beyond the scope of this paper to enter into the very complicate problem
of looking for
the suitable models for the collisionless stellar systems by an analysis in the phase space
and a research 
of the $DF$ which maximizes the (\ref{entr}), or to examine the thermodynamic 
properties of the family models which are able to explain the features of partially 
relaxed anisotropic stellar systems (see, e.g., Stiavelli \& Bertin, 1987, Bertin
\& Trenti, 2003, and references therein).
Nevertheless, our limited contribution to the wide discussion existing in the literature will be
to underline as in a 
stellar component, 
embedded 
in a second dark 
matter subsystem (as realistically thought, e.g., Ciotti, 1999, and references therein), 
a truncation is spontaneously introduced in coordinate space, due to the presence of 
a scale length induced
from the dark halo, as long as virial equilibrium holds.
That is the {\it tidal radius} which we discovered has to exist when two-component
models are considered with two different power-law density distributions and two inner 
homogeneous cores
(Secco, 2000; Secco, 2001, hereafter LS1), under some constraints on the exponents.

The consequence of the existence of a special
configuration characterized by this {\it tidal radius}
are analyzed here.

Finally, we will gain more insight into the physical meaning of the special 
configuration considered by introducing the {\it thermodynamic information} quantity
(Layzer, 1976).

Even if some considerations which follow are more general and may also be
extended to spirals, we will limit our considerations 
to the collisionless stellar systems, as the ellipticals are considered.

\section{Looking for a special virial configuration}
To introduce the problem in a general way,
we start by considering the potential well of a given spherical virialized dark matter
(hereafter, $DM$) halo
of mass $M_D$ and virial radius $a_D$, with a 
density radial profile as follows:
\begin{equation}
\label {nfw}
\rho(r)=\frac{\rho_o}{(r/r_o)^{\gamma}[1+(r/r_o)^{\alpha}]^{\delta}}~~~~
;~\delta=(\beta-\gamma)/\alpha
\end{equation}
where $\rho_o$ and $r_o$ are its
characteristic density and  its scale radius, respectively.
These kinds of profiles have already been introduced by Zhao (1996) and by 
Kravtsov et al.(1998) in order to generalize the universal profile
proposed by Navarro, Frenk \& White (hereafter, NFW) (Navarro et al.1996,
Navarro et al. 1997)which is obtained from eq.(\ref{nfw})
as soon as ($\alpha=1; \beta=3;\gamma=1;\delta=2$). Hereafter, we will name them   
{\it Zhao profiles}. 

The question which arises is the following:  
{\it Does a special virial configuration exist among the infinitive number of a priori
possible virial configurations which the luminous (Baryonic) component (B) may assume inside the 
given dark one (D)?}

\begin{figure}[!h]
\begin{center}
\includegraphics[height=11cm,width=15cm]{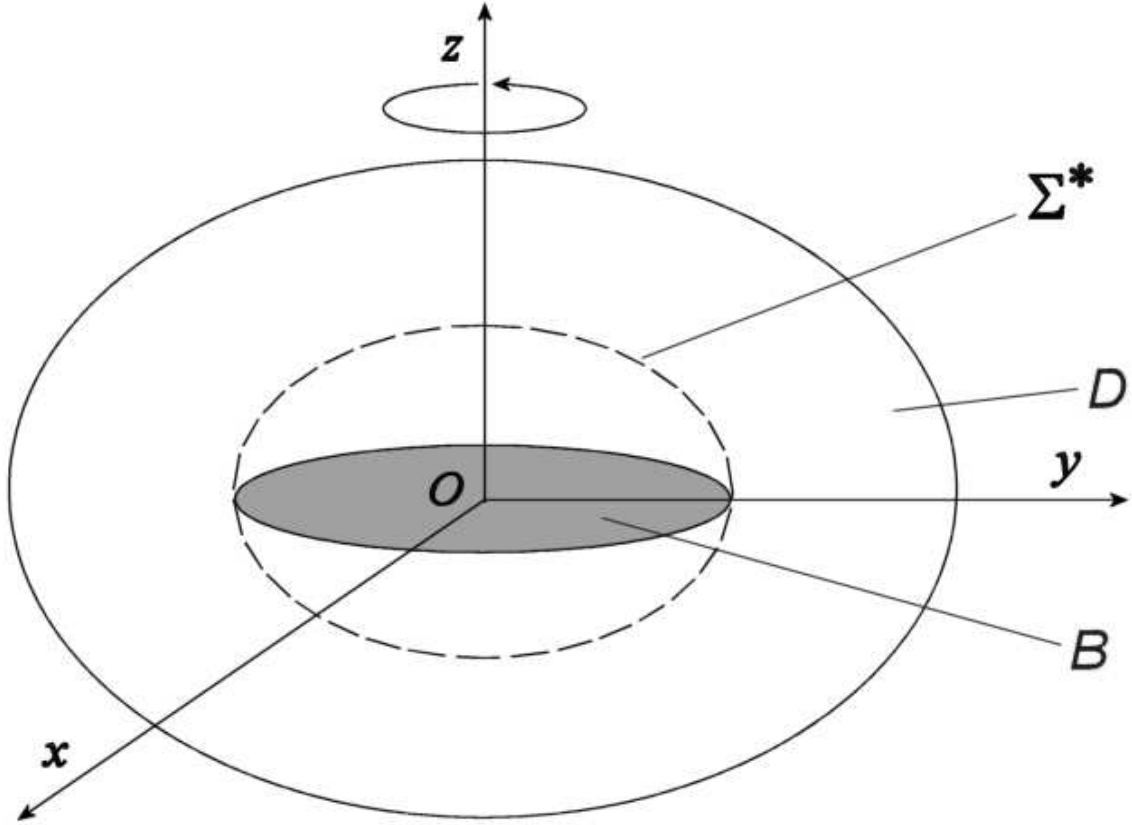}
\caption{Two-component model: $D$ is the dark matter ($DM$) halo spheroidal component, 
$B$ is the bright (Baryonic) inner one. Here the $B$ component is non-homothetic to 
the $D$ one instead of what occurs in  
this paper. Fig.1 shows a $B$ spheroid with an axis ratio smaller than
that of the $D$ one in order to show what is, in the general case,
the dark matter fraction which, according
to Newton's first theorem, exerts dynamical effect on the embedded $B$ subsystem. 
In this general case the dark matter fraction is inside the surface $\Sigma^*$ which,
in the 
homothetic case, coincides with the $B$ contour 
(Raffaele, 2003).}
\label{potenz}
\end{center}
\end{figure}

\subsection{Tensor virial formalism}
In order to find the answer we need to use the tensor virial theorem extended
to two components: $D+B$ (Brosche et al. 1983; Caimmi et al. 1984; Caimmi \& Secco, 1992).
In a two-component system in which one (B) (Baryonic or Bright; in this context also: stellar) is completely embedded in 
the other (D) (of $DM$) and 
each of them is
penetrated by the other, the following tensor virial stationary equations hold:
\begin{equation}
\label{virial}
2(T_u)_{ij} =-(V_u)_{ij};~~(u=B,D;~~ i,j=x,y,z)
\end{equation}
where $(T_u)_{ij}$ is the kinetic-energy tensor and $(V_u)_{ij}$ is the 
Clausius' virial 
tensor which  
splits into two terms: the 
self potential-energy tensor, $(\Omega_u)_{ij}$, and 
the tidal potential-energy tensor, $(V_{uv})_{ij}~~ (u,v=B,D)$, 
due to the gravitational force which the $v$ subsystem 
exerts on the $u$ one.
The eqs.(\ref{virial}) yield the following pair of tensor equations:
\begin{eqnarray}
\label{eq:vir2}
2(T_B)_{ij} =- (\Omega_B)_{ij} -(V_{BD})_{ij}\\
2(T_D)_{ij} =- (\Omega_D)_{ij} -(V_{DB})_{ij}
\end{eqnarray}
 Therefore, e.g., in the case of the inner $B$ component, we have: 
\begin{equation}
\label{eq:VBAij}
(V_{BD})_{ij} = \int\rho_B x_i\frac{\partial\Phi_D}
{\partial x_j}{\rm\,d}\vec{x}_B~; 
\end{equation} 
where $\Phi_D$ is the gravitational potential due to 
the mass distribution of the $D$ component. 

\begin{figure}[!h]
\begin{center}
\includegraphics[height=15cm,width=12cm,angle=-90]{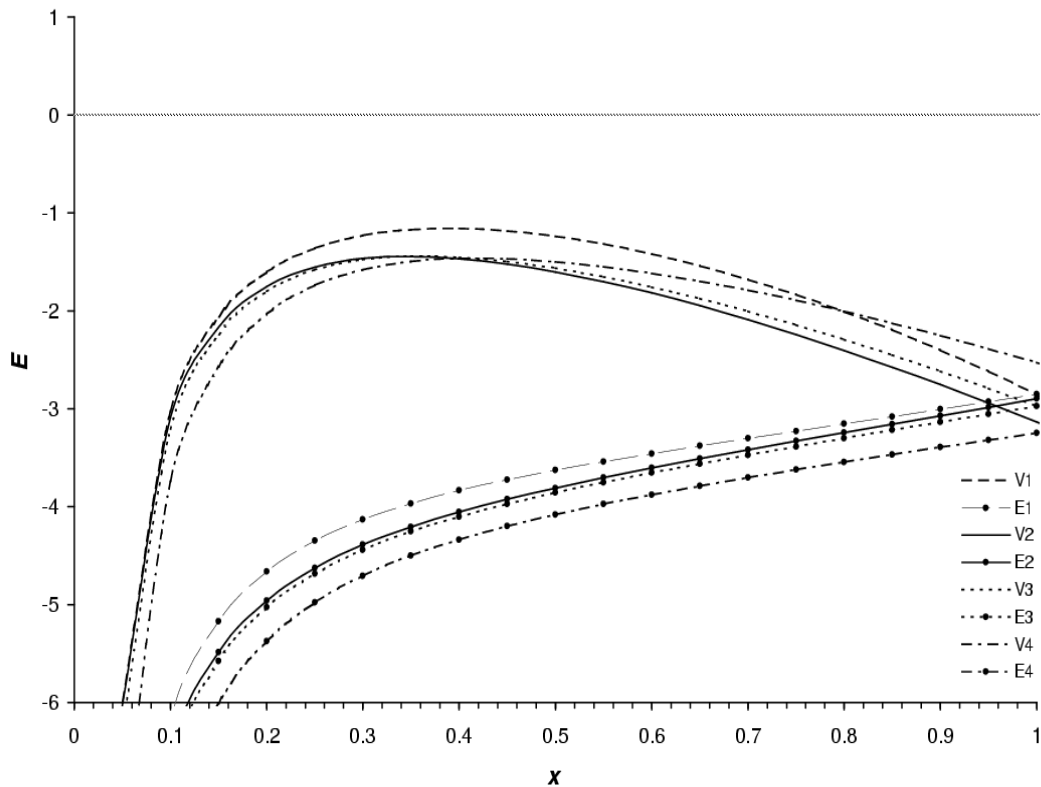}
\caption{The energy trends of the $B$-system as a function 
of size ratio of Baryonic to $DM$ components, $x=a_B/a_D$, normalized at the factor $(GM_B^2F)/a_D$. The $V$-curves 
represent the Clausius' virial 
energies, the $E$-curves the corresponding total potential 
energies, in the cases of Tab.1 (Raffaele, 2003).}

\label{potenz}
\end{center}
\end{figure}

Even if we do not look for the distribution functions which correspond
to the models considered in the next subsection, it may be useful to
remember the link between the tensor virial quantities and the phase space, given 
by considering 
the moment equations of the second order, in the coordinate space, of the $II^o$ Jeans 
equations which, in turn, are
the second
order moments, in the velocity space, of the Boltzmann equation (Binney \& Tremaine, 1987, 
Chap.4; Chandrasekhar, 1969, Chapt.2).
It follows that in a system of collisionless particles
with istantaneous $\vec{x}$
position and $\vec{v}$ velocity,
characterized by a distribution function $f(\vec{x},\vec{v})$, with spatial 
density in phase 
space, $\nu(\vec{x})= \int
f(\vec{x},\vec{v}){\rm\,d}\vec{v},$ and mass density 
$\rho(\vec{x})= m_*~\nu(\vec{x})$ ($m_*$ being the average stellar mass), 
the kinetic-energy tensors $T_{ij}$, $T^\prime_{ij}$, $\Pi_{ij}$, 
are defined as:
\begin{eqnarray}
\label{eq:Tij}
T_{ij}=\frac12\int\rho\overline{v_iv_j}{\rm\,d}\vec{x}=
T^\prime_{ij}+\frac12\Pi_{ij}~; \\
\label{eq:T'ij}
T^\prime_{ij}=\frac12\int\rho\overline{v}_i\overline
{v}_j{\rm\,d}\vec{x}~;~\Pi_{ij}=\int\rho\sigma_{ij}^2{\rm\,d}\vec{x}~;
\end{eqnarray}
that is by the mean of square 
velocities ($\overline 
{v_iv_j}$), by the square of mean velocities ($\overline 
{v}_i\overline{v}_j$) of streaming motions, and by the random square
velocity components ($\sigma_{ij}^2$), respectively.
The general expression for the mean $\overline{v_iv_j}$ is as usual:
\begin{equation}
\overline{v_iv_j}  =\frac1\nu\int f v_iv_j{\rm\,d}\vec{x}
\end{equation}
In the case of one single component, the Clausius' virial tensor, $V_{ij}$,
matches the self-potential energy tensor, $\Omega_{ij}$, that is:
\begin{equation}
V_{ij}= \Omega_{ij}=-\frac12\int\rho\Phi_{ij}{\rm\,d}\vec{x}
\end{equation}
$\Phi_{ij}$ is the tensor potential (e.g., Chandrasekhar, 1969) defined 
as: $$\Phi_{ij}= G \int\rho(\vec{x^{\prime}}) \frac
{(x_i-x_i^{\prime})( (x_j-x_j^{\prime})}{\mid {\vec{x}-\vec{x^{\prime}}\mid}^3}
{\rm\,d}\vec{x}$$
with $G$ the gravitational constant. 
It should be underlined that, in general, the presence of one external 
component causes the 
non-equality between the 
total potential energy tensors
of the subsystems and their Clausius' virial tensors, as occurs in 
the case of
the single 
component (see, e.g., LS1). Moreover, we should remember that
Clausius' tensors are made only by forces and positions and then
the mass fraction of the dark outer component which enters
in the $B$ Clausius' tensor is only that which exerts dynamic effects on $B$,
according to Newton's first theorem.

\subsection{Two-component models}

Then we need to model the two components (Fig.1). 
The models we consider here are the same as in LS1. That is they  are 
built up of two homothetic similar strata spheroids  
with two power-laws
and two different homogeneous cores in the central regions.
It should be noted that they correspond, to a good extent,
\footnote{
Generally, the deviations are only of a few per cent and less than the $10\%$ related
to the Clausius' minimum location and to its value, respectively.}
to deal 
with two spheroids with smoothed 
profiles of this kind (spherical case\footnote{Even if the considerations which follow are more general, for the 
sake of 
simplicity, we  will often limit ourselves to the spherical case
without losing the validity of spheroidal case, which may be recovered simply by introducing
a form factor, $F$, depending on the axis ratio (see, eq.(\ref{Avbd})).}):

\begin{eqnarray}
\label{profiD}
\rho_D=\fr{\rho_{oD}}{1+(\fr{r}{r_{oD}})^d};~ C_D=\fr{a_D}{r_{oD}}\\
\label{profiB}
\rho_B=\fr{\rho_{oB}}{1+(\fr{r}{r_{oB}})^b};~ C_B=\fr{a_B}{r_{oB}}
\end{eqnarray}
$C_B$ and 
$C_D$ are the two concentrations of the two components. In the LS1 models,
$r_{oB}$ and $r_{oD}$ were the radii of the two different homogeneous cores, which
typically assumed one tenth of the virial radii, $a_B$ and $a_D$, respectively.
According to these, the concentrations in the 
smoothed profiles both become equal to ten.
 The density profiles of kind (\ref{profiD}) and (\ref{profiB}), have
 the advantage that they may be considered as a generalization of pseudo-isothermal profiles
which, in turn, may be regarded as sub-cases of the 
more general {\it Zhao profiles} when: $\gamma=0; \beta=\alpha=b,d$.

But a realistic elliptical model has to be: e.g., a stellar component with a
Hernquist (1990) (hereafter, Her) density profile and a dark halo with a {\it cored} or {\it non-cored}
NFW profile (that means, according to
eq.(\ref{nfw}), respectively: $\alpha=1; \beta=4;\gamma=1;\delta=3$ and
 $\alpha=1; \beta=3;\gamma=0;\delta=3$; for the {\it cored} NFW,
$\alpha=1; \beta=3;\gamma=1;\delta=2$, for the {\it non-cored} NFW) (as in Marmo, 2003, where two
homeoidally striated ellipsoids are considered).

Then, the problem of transfering the outputs obtained with two cored power-law profiles
(which also hold, to a good
extent,
for the models with smoothed profiles (\ref{profiD}), (\ref{profiB})) to the more general 
class of models
with {\it Zhao profiles} given by eq.(\ref{nfw}), is still open, even if some preliminary
considerations will
be made in the subsect. 4.5.

Our aim is to try to explain some scaling relations for elliptical galaxies,
which  are essentially
relationships among the exponents of the 
three
quantities:

$r_e=$the effective radius,

$I_e=\frac{L}{2\pi r_e^2}=$mean effective surface brightness within 
$r_e$
		
$\sigma_o$=the central projected velocity dispersion.

The advantage of a simple {\it cored} power-law model is that it is able to extract,
in a completely 
analytical way, some of these main correlations, highlighting the interplay of the parameters. 
Moreover,
this preliminary analysis may also underline what are simply details in the model
and what, on the contrary, is stictly connected
with the physical reason for the existence of a FP for two-component virialized systems.
That allows us to open the road
for a generalization of the present results.

\section{The special virial configuration}
As shown in the previous papers, a {\it special configuration} for the $B$ component
arises inside these kind of homothetic (see, LS1) models (for the non-homothetic, see, 
Secco, 2000), which describe its evolutive
pattern, obtained by contraction inside a $D$ one, under the assumption that $D$ is 
at fixed size and shape.
For the sake of simplicity, we will
assume that the outer component is frozen,  without considering this constraint
to be too essential in order to determine the main features of the dynamic
evolution we are dealing with ( see, LS1). The main reason for this assumption is indeed
that the masses of
the two components are not equal, the outer one being about ten times the inner one. As a
consequence, tidal 
influences between the subsystems are not symmetric; the one acting from
inner to the outer
is actually
weaker than the reverse
(Caimmi \& Secco 1992).
Moreover, even if a contraction effect is
induced by inner density distribution on the inner regions of
outer halo, during the dynamical evolution, as already underlined by 
Barnes \& White (1984) , that effect  
does not cause  
a dramatic modification of the outer mass distribution, if there is 
a supernovae- driven outflow, according to 
N-body simulations (Lia 
et al. 2000). On the other side, models in which this constraint has been changed with
some less stringent additional conditions (Caimmi, 1994), seem to come to the same conclusion.

The special configuration appears because a maximum in the Clausius virial
energy trend, as $B$ contracts inside $D$, (and then a minimum in the kinetic energy) 
exists under the following constraints on the exponents:
\begin{equation}
\label{constr}
0\le b <~3~;~ 0\le d <~2 \Rightarrow (b+d)< 5
\end{equation}
The total potential energy, $(E_{pot})_B$, on the contrary  
is always
monothonic (Fig.2). Indeed, by definition:

\begin{equation}
(E_{pot})_B=\Omega_B+W_{BD}
\end{equation}
where $\Omega_B$ is the self-potential energy tensor trace
and $W_{BD}$ is
the interaction-energy trace of the tensor defined as:
\begin{equation}
\label{eq:WBAij}
(W_{BD})_{ij} = -\frac12\int_B\rho_B (\Phi_D)_{ij}
{\rm\,d}\vec{x}_B~; 
\end{equation} 
As already underlined,
in general, $(W_{BD})_{ij}$ does not match $V_{ij}$ (see, Caimmi \& Secco, 1992).

\begin{table}
\caption{Physical parameters of the models considered in the figures $2,~5,~6$. 
The common density profile
parameters (eqs.(\ref{profiD}, \ref{profiB})) are: $C_B=C_D=10,~m=8.5$. Moreover
$x_t=a_t/a_D$; for the definition of $\nu_{\Omega B};~\nu_{\Omega D};~\nu'_V$, see text.}
\vspace{0.3cm}
\begin{center}
\footnotesize
\begin{tabular}{ccccccc}
\hline
&&&&&&\\
cases &$b$&$d$&$x_t$&$\nu_{\Omega B}$&$\nu_{\Omega D}$&$\nu'_V$\\
&&&&&&\\
\hline
&&&&&&\\
1)&$0.0$&$0.0$&$0.389$&$0.300$&$0.300$&$0.300$\\

2)&$0.0$&$0.5$&$0.346$&$0.300$&$0.312$&$0.333$\\
3)&$0.5$&$0.5$&$0.361$&$0.312$&$0.312$&$0.313$\\
4)&$1.5$&$0.5$&$0.419$&$0.367$&$0.312$&$0.254$\\
&&&&&&\\
\hline
&&&&&&\\

\hline

\end{tabular}
\end{center}
\label{tab1B}
\end{table}

By referring to the LS1-models with profiles of kind (\ref{profiD}),(\ref{profiB}), the 
special configuration (which 
appears under the constraints
on the exponents of the density power-laws and by taking a frozen
$D$ subsystem, as seen in the same paper), corresponds to the 
dimension of the $B$ component characterized by the following semimajor
axis, called {\it tidal radius}:
\begin{equation}
\label{at}
a_t=\Big( \frac{\nu_{\Omega B}}{\nu'_V}
\frac{1}{(2-d)}\frac{M_B}{M_D}\Big)^{\frac{1}{3-d}}~a_D 
\end{equation}
where $M$ and $a$ are, respectively, the mass and the major semiaxis
of the subsystem considered. 
The coefficient $\nu_{\Omega B}$ enters the integral which
gives the self- potential energy tensor of the $B$ component (see, LS1),
so it is a function only of the $b$ exponent; the coefficient $\nu'_V$
is defined in the following way:

\begin{equation}
\label{nupv}
\nu'_V\simeq\frac{9}{2}[(\nu_B)_M(\nu_D)_M]^{-1} 
 \frac{C_B^{-(b+d)}}{(3-d)[5-(b+d)]}
\end{equation}
where
$(\nu_B)_M, (\nu_D)_M$ are defined in LS1.
$\nu'_V$, which is a function of $b$ and $d$ and of the two 
concentrations $C_B=C_D$, enters in the definitions of the tidal tensor trace:
\begin{equation}
\label{Avbd}
V_{BD}\simeq- \nu'_V~G \frac{M_B \widetilde{M_D}}{a_B}F;
\end{equation}
where $F$ is the form factor (see, Marmo \& Secco,  2003, hereafter MS3) and $\widetilde{M_D}$ is the fraction of $D$ matter 
exerting
dynamical effect on $B$, according to Newton's first theorem. To a good extent
it is given by:
\begin{equation}
\label{Mtilde}
\widetilde{M_D}= M_D (\frac{a_B}{a_D})^{3-d}
\end{equation}
The same mass fraction normalized to $M_B$ becomes:
\begin{equation}
\label{mtilde}
\widetilde {m}=\frac{M_D}{M_B}(\frac{a_B}{a_D})^{3-d}
\end{equation}
Moreover we define the total mass inside the B-structure as:
\begin{equation}
\label{mass}
M^*_{tot}= M_B+\widetilde{M_D}=M_B(1+\widetilde{m})
\end{equation}

These quantities also enter into the definition of
the 
$B$-Clausius' virial tensor trace as follows:
\begin{equation}
\label{Cla}
 V_B \simeq \Big[-\nu_{\Omega B}\frac{GM_B^2}{a_B}F- \nu'_V 
		  \frac{GM_B\widetilde{M_D}}{a_B}F \Big]
\end{equation}
or in the form normalized  by the factor $\frac{ a_D}{GM_B^2F }$:
\begin{equation}
\label{virn}
\widetilde{V_B}\simeq-\frac{\nu_{\Omega B}}
		   {x}-\nu'_V ~m~x^{2-d};~~x=\frac{a_B}{a_D};~~m=\frac{M_D}{M_B}
\end{equation}

Why this configuration of the $B$ component inside the fixed dark matter potential
well is so special, may be understood by considering, in the next section, 
its mechanical and thermodynamical 
properties.

\section{Special configuration: mechanical and thermodynamical properties}
\label{MTpro}
\subsection{Mechanical arguments}
In order to understand the full meaning of the {\it tidal radius}, which is defined
by eq.(\ref{at}), and to be able to consider the
thermodynamical processes of the $B$ component, we have to look at the 
physics related to Clausius' virial energy, $V_B$.
By definition, the Clausius' virial tensor trace is given by:
\begin{eqnarray}\
\label{Clau}
V_B= \Omega_B + V_{BD}\\
\label{omB}
\Omega_B= \int\rho_B\sum_{r={1}}^{3}~ x_r\frac{\partial\Phi_B}
{\partial x_r}{\rm\,d}\vec{x_B}~= \int\rho_B
 (\vec{r_B}\cdot\vec{f_B}){\rm\,d}\vec{x_B}\\
\label{VBA}
V_{BD}= \int\rho_B\sum_{r={1}}^{3} x_r\frac{\partial\Phi_D}
{\partial x_r}{\rm\,d}\vec{x_B}~=\int\rho_B(\vec{r_B}\cdot\vec{f_D})
{\rm\,d}\vec{x_B};
\end{eqnarray}
where $\vec{f_B}$ and $\vec{f_D}$ is the force per unit of bright mass due to 
self gravity and the dark matter gravity at the
point $\vec{r}_B$, respectively.
By definition, the work done by the self gravity forces, $L_s$, in order to assemble the $B$-elements
from the infinity is given by the self- potential energy $\Omega_B$ and the work done
by the tidal
gravity forces, $L_t$, in order to put the $B$ component together with the $D$ one from 
infinity
through all the tidal distorsions (see, Caimmi \& Secco, 2004), is given by the
tidal potential energy $V_{BD}$. 

Then a small variation $\delta V_B$ for a small displacement 
$\delta\vec{ r}_B$ of all $B$ points,
has the following meaning:   
\begin{equation}
\label{lavtot}
\delta V_B=\delta L_s+\delta L_t
\end{equation}

\subsection{Small departures from virial equilibrium}

We will now consider what is the mathematical form of the $V_B$ potential energy variation
as soon as the $B$ 
inner system contracts or expands
its initial volume $S_o$ of a small quantity $\Delta S_o$.
Following Chandrasekhar's analysis (Chandrasekhar, 1969, Chapter 2) the following holds:
by definition, the Clausius virial is a global, integral parameter of an {\it extrinsic} 
attribute, 
that means of a quantity which is not intrinsic to the fluid element (like pressure or 
density) but something, we name $F(\vec{x})$, which it assumes simply by virtue of its 
location such as the 
gravitational potential and its first derivative. The variation of the
integral:
\begin{equation}
\delta \int_{S_o}\rho_B F{\rm\,d}\vec{x}_B=\int_{S_o+\Delta S_o}\rho_B F{\rm\,d}
\vec{x}_B-\int_{S_o}\rho_B F{\rm\,d}\vec{x}_B
\end{equation}
when the istantaneously occupied volume by the fluid changes from 
$S_o$ to $S_o+\Delta S_o$ by subjecting its boundary to the displacement 
$\vec{\xi}(\vec{x},t)=\vec{x}-\vec{x}_o$,
may be transformed into the integral over the unperturbed volume, that is:
\begin{equation}
\delta \int_{S_o}\rho_B F{\rm\,d}\vec{x}_B=\int_{S_o}\rho_B \Delta F{\rm\,d}\vec{x}_B
\end{equation}
where $\Delta F$ is the Lagrangian change in $F$ consequent to the displacement $\vec{\xi}$.

The extension of this analysis to two-component systems has been performed (Caimmi \& Secco,
2004) with the following result:

\begin{eqnarray}
\label{varOM}
\delta \Omega_B=-\int_{S_o}\rho_B\sum_{r={1}}^{3}~ \xi_r\frac{\partial\Phi_B}
{\partial x_r}{\rm\,d}\vec{x}_B\\
\label{varTI}
\nonumber
\delta V_{BD}= -\int_{S_o}\rho_B\sum_{k={1}}^{3}\sum_{r={1}}^{3}\xi_k\frac{\partial}
{\partial x_k}(x_r\frac{\partial\Phi_D}
{\partial x_r}){\rm\,d}\vec{x}_B-\int_{M_o}\rho_D\sum_{r={1}}^{3}~ \xi^{\prime}_r\frac{\partial\Phi_B}
{\partial x_r}{\rm\,d}\vec{x}_D\\
+\int_{M_o}\rho_D\sum_{k={1}}^{3}\sum_{r={1}}^{3}\xi'_k\frac{\partial}
{\partial x_k}(x_r\frac{\partial\Phi_B}{\partial x_r}){\rm\,d}\vec{x}_D
\end{eqnarray}
where the unperturbed volume of $D$-component is $M_o$ and $\vec{\xi^{\prime}}(\vec{x},t)$ is the amount
of the perturbation in the point domain of the same component.

Under the assumption of a frozen dark component, $\vec{\xi'}$ vanishes and
the main result holds:

\begin{eqnarray}
\nonumber
\delta V_B=\delta L_s+\delta L_t\simeq-\int_{S_o}\rho_B\sum_{r={1}}^{3}~ \xi_r\frac{\partial\Phi_B}
{\partial x_r}{\rm\,d}\vec{x}_B\\
-\int_{S_o}\rho_B\sum_{k={1}}^{3}\sum_{r={1}}^{3}\xi_k\frac{\partial}
{\partial x_k}(x_r\frac{\partial\Phi_D}
{\partial x_r}){\rm\,d}\vec{x}_B=(\delta V_B)_{a_D}
\end{eqnarray}
By definition of the {\it tidal radius}, which is the $B$ dimension at the
maximum of its Clausius virial energy, at frozen $a_D$, the $(\delta V_B)_{a_D}$ is 
stationary at $a_t$ (Fig.2)(see, LS1). Therefore, by moving of a 
virtual\footnote{Here virtual means at frozen D.} displacement $\delta a_B$,
from $a_B=a_t$, we have:
\begin{equation}
\label{lavts}
\delta L_s+\delta L_t\simeq(\delta V_B(a_t))_{a_D}=0
\end{equation}
This means that the configuration at $a_t$ satisfys the d'Alembert Principle
of {\it virtual works} (see also LS1).
The physical reason is the following: if, e.g., $B$ contracts, less
dark matter enters inside the $S_B$ surface, in the 
meanwhile the self gravity increases. The opposite occurs if $B$ expands itself.
Therefore, even if both forces are attractive, the works
which correspond to them, for a virtual 
displacement, are of opposite signs (see, LS1). Therefore, the tidal radius 
configuration is an 
equilibrium configuration even if not stable because the total potential energy
of $B$ has not a minimum (Fig.2).

The consequence of these mechanical arguments with the property of 
the {\it tidal configuration} to be able to distribute in about the equal
parts the self- and the tidal- energies (see, next subsection and LS1), is to yield some 
outputs
which are in good agreement with the corresponding observable scaling relations
related to the elliptical galaxies, and in general to the existence of a FP for two-component
virialized systems, as
we have already highlighted in the past papers 
(LS1, MS3).

\subsection{Scaling relations at the special configuration}
Even if the physical explanation for the main features of FP for virialized structures
and in particular for the dynamically hot ellipticals ( Bender et al., 1992), is still an 
open question, the two main roads which are present in the literature (see, e.g.,
Renzini \& Ciotti, 1993; Ciotti et al., 1996; Bertin et al., 2002) may be summarized as:

a) The {\it tilt} of the FP is due to the stellar population effect.

b) The {\it tilt} is due to a non-homologous structural effect. The light profile 
is not an universal de Vaucoleurs profile but a Sersic profile (1968) which changes with 
the galaxy luminosity: the Sersic index $n$ increases as the luminosity increases
({\it weak homology}, Bertin et al., 2002).

Neither a) nor b) seem to give the proper answer to the scaling relations problem;
the population effect disappears in the K-band (Maraston, 1999), contrary to what is observed
(Pahre et al., 1998). Therefore, a metallicity sequence of an old stellar population
may at most fit the trend observed in the B-band (Gerhard et al., 2001).
Moreover with the {\it weak homology} the observed {\it tightness} of the FP appears to be hard
to explain (Bertin et al., 2002).

According to theory performed in the papers: Secco (2000), LS1, MS3, we tried to 
explain the {\it tilt} by assuming a strict
homology which does not imply a constant ratio $M_B/L$ due to the presence of a dynamical 
effect caused by the scale length induced on the gravitational baryonic field from the
dark matter halo distribution.

We come back to the virial equations (\ref {virial}), in the trace form, for a B-component 
completely embedded in a dark halo.
The FP we obtain (see, LS1):
\begin{eqnarray}
\label{FPII*}
r_e=c_2c_1^{-1}\sigma_o^A~I_e^{B}G_2~\frac{L}{M^*_{tot}}; ~A=2;~ B=-1\\
\label{g2}
G_2=
\frac{1+\widetilde{m}}{F[\nu_{\Omega B}+\widetilde{m}\nu'_V]}
\end{eqnarray}
seems, at a first sight, not substantially changed  in respect to the one we get by using the
virial equations for a single component system.

But if we consider the trace of eq.(\ref{eq:vir2}), at the special configuration with the condition
that the bright component is pressure supported (i.e., peculiar kinetic energy
$T_{pec}$ is dominant with respect to the rotational kinetic energy $T_{rot}$) we obtain:

 \begin{equation}
 \label{fuprox1}
 \frac{1}{2}M_B<\sigma^2>\simeq \left(\frac{-\Omega_B-V_{BD}}
 {2}\right)_{a_B=a_t};~~T_{rot}<<T_{pec}
 \end{equation}
where $<\sigma^2>$ is the mean square velocity dispersion of the stars.
By adding at $a_t$ the equipartition between the {\it self- and tidal energy}. 
the previous equation becomes:

\begin{equation}
\label{approx}
a_t\simeq \left( \frac{\frac{1}{2}M_B\frac{\sigma_o^2}{k_v} a_D^{3-d}}
{\nu'_V GM_B M_D F}\right)^{\frac{1}{2-d}}
\end{equation}
which, instead of eq.(\ref{FPII*}) yields the following FP:

\begin{equation}
\label{FPN}
r_e\sim \sigma_o^{\frac{2}{2-d}}a_D^{\frac{3-d}{2-d}}m^{-\frac{1}{2-d}} M_B^{-\frac{1}{2-d}}
\end{equation}
That means:

\begin{eqnarray}
\left\{
\label{guess}
\begin{array}{l}
\sigma_o^A\equiv \sigma_o^{\frac{2}{2-d}}\\
\label{guesI}
I_e^B\sim a_D^{\frac{3-d}{2-d}}m^{-\frac{1}{2-d}} M_B^{-\frac{1}{2-d}}
\end{array}
\right.
\end{eqnarray}

\subsection{Output vs. observables}

I) From the first equality we have:

\begin{equation}
A=\frac{2}{2-d}
\end{equation}

II) Moreover, 
due to the two relationships which connect the FP coefficients, $A$,$B$ with the 
observed {\it tilt} (see, e.g., Djorgovski \& Santiago, 1993), that is:
\begin{eqnarray}
\left\{
\begin{array}{l}
A=\frac{2(1-\alpha_t)}{1+\alpha_t}\\
B=-\frac{1}{1+\alpha_t}
\end{array}
\right.
\end{eqnarray}
we immediately obtain
the {\it tilt} of the FP:
\begin{equation}
\alpha_t=\frac{1-d}{3-d}
\end{equation}
and the other coefficient:
\begin{equation}
B=-\frac{3-d}{2(2-d)}
\end{equation}
only as functions of the dark matter distribution.
First of all it shoud be noted that the quantities $A,B,\alpha_t$ which define the 
FP and its {\it tilt} are independent of $m$. That means the galaxies which belong to the FP
may have a different fraction of baryonic matter in respect to the dark one, but their 
dark matter density profile must be the same.
To probe the issue of the first identity 
we choose, e.g.:

$d=0.5 \Rightarrow$ $A=1.33$, $B=-0.83$ and $\alpha_t=0.20$

in good agreement with the observations in the $B-$band.

In the range: $d=0\div1\Longrightarrow~A=1\div2~;-B=~0.75\div1;~\alpha_t=0.33\div0$
in agreement with the data related to all bands (Marmo, 2003, Tab.1.1 and the references therein).

This allows us to expect (see, LS1) that
in other families of galaxies with dark matter halos of the same
kind of elliptical galaxies, as, e.g., the spirals, a Fundamental Plane also has 
to exist which has the same $A$, $B$ and $\alpha$ exponents, totally independent
of a completely different luminous mass distribution.
This is in good agreement with the universal FP discovered
by Burstein et al. (1997).

III) It should be noted that the values of  $A,~B,~\alpha_t$ which define the FP as a whole, 
are not directly linked with
the past cosmological conditions, but they are only a function 
of the dark matter distribution $d$. This is in agreement with what Djorgovski
already noted (1992) on the basis of Gott \& Rees (1975), Gunn (1987), 
Coles \& Lucchin (1995), occurs for
the scaling
relations in a CDM scenario.  Indeed, from a cosmological point of view the FP means
the following relationship (Djorgovsky, 1992), which we also recover in LS1:

$$2n_{rec}+10=A(1-n_{rec})-B(12\alpha_t+4n_{rec}+8),$$

$n_{rec}$= effective spectral index of perturbations.

IV) But in the projections of the FP on the coordinate planes the dependence on the 
cosmological spectral index appears
via the parameter, $\gamma'$, we introduced in LS1, as:
\begin{equation}
\label{gam'}
\frac{1}{\gamma'(M)}= \frac{1+3\alpha_{rec}(M)}{3}=\frac{5+n_{rec}}{6}
\end{equation}
where, according to
Gott \& Rees (1975b) and Coles \& Lucchin (1995, Chapts.14, 15 ), $\alpha_{rec}$ is the 
local slope of the CDM mass variance, $\sigma^2_M$, at recombination time $t_{rec}$, 
given by:

$$\alpha_{rec}=-\frac{dln~\sigma_M (t_{rec})}{dlnM},$$

If the total energy of the system is conserved during the transition from a 
maximum expansion phase to the virialization, with its re-distribution among the collisionless
ingredients
by a violent relaxation mechanism, the following dependences on $m$ and $M_B$,
for the three main quantities of FP, hold:

\begin{eqnarray}
\label{re}
r_e\sim m^rM_B^R~;~~
r=\frac{(3-d)-\gamma'}{\gamma'(3-d)};~R=1/\gamma'\\
\label{ie}
I_e\sim m^i M_B^I~;
~~I=i=2\frac{\gamma'-(3-d)}{\gamma'(3-d)}\\
\label{sig}
\sigma_o\sim m^sM^S_B~;
s=-\frac{1}{2}\frac{(3-d)-\gamma'}{\gamma'(3-d)};~S=\frac{1}{2}
\frac{\gamma'-1}{\gamma'}
\end{eqnarray}

Therefore, the projected scaling relation, such as the Faber-Jackson relation (hereafter FJ),
turns out to be:
\begin{equation}
\label{FJII}
L\sim m^{2\frac{(3-d)-\gamma'}{(3-d)^2(\gamma'-1)}}
~\sigma_o^{\frac{4\gamma'}{(\gamma'-1)(3-d)}}
\end{equation}
not only directly related to the dark matter distribution, via the exponent 
$d$, but also to the perturbation spectrum, via $\gamma'$.

For a typical galaxy dark matter halo of $M_D\simeq 10^{11}M_{\odot}$, it turns out that $\gamma'=2$
(Gunn, 1987).
If $d=0.5$, we obtain:

$$L\sim M_B^{0.8};~~ r_e\sim M_B^{0.5}~\Rightarrow~~ 
I_e=L/2\pi r_e^2\sim~ M_B^{-0.2}$$

It should be noted that $I_e$ decreases as $M_B$ increases as soon as:
$$3-d~>~\gamma';$$
that is $d~<~1$.
On this mass scale, the FJ becomes:

$$L\sim m^{0.16}\sigma_o^{3.2}$$.

If $d=1$, then $\alpha_t=0$ and as a consequence: $$L\sim \sigma_o^4$$ without dependence on $m$.

V) Moreover on the Clausius' minimum we have:

$$\left(M^*_{tot}\right)_t=M_B\Big (1+\frac{\nu_{\Omega B}}{\nu'_V (2-d)}\Big)$$

If $b$ and $d$ are universal, then the consequence is the proportionality of the two masses
and then of the two ratios:
\begin{equation} 
\label{lsum}
L/(M^*_{tot})_t\sim~~L/M_B 
\end{equation}

VI) From an observational point of view, we know (Cappellari et al., 2004) that,
for a sample of E and S0, either fast rotators or nonrotating giant ellipticals, the 
following tight
correlation holds:
\begin{equation}
M/L\sim \sigma_o^{0.8}
\end{equation}
From the previous scaling relations, we obtain:
\begin{equation}
\label{mlsig}
M_B/L\sim\sigma_o^{ 
\frac{2\alpha_t \gamma'(M_D)}{\gamma'(M_D)-1}}
\end{equation}
which on this dark matter scale yields an exponent of $\sigma_o$
exactly equal to $0.8$
without distinguishing between $L/(M^*_{tot})_t$ and $L/M_B$,
according to eq.(\ref{lsum}). 
That also has to be compared with
J\o rgensen's (1999) value:
$0.76\pm 0.08$.
It should to be noted, as in the relationship (\ref{mlsig}), that the dependence on the other 
factor $m$ is completely negligible. Indeed it turns to be:~$m^{0.04}$.

VII) Another issue is:
the ratio of the dark matter fraction over total
mass inside the bright radius $a_B$.

At $a_t$, it becomes:

$$\left (\frac{\widetilde{M_D}}{M^*_{tot}}\right )_t=\frac{1}{1+\frac{\nu'_V}
{\nu_{\Omega B}}
(2-d)}$$

That means that it only depends on the luminous and dark density 
profiles. If they are both universal for the galaxy family considered, this dark matter fraction has to be 
the same
for all members.

It should be noted that this result is
independent of the total mass ratio, dark over bright, $m$.

If the most probable value for $d$ is around $0.5$
and $b$ ranges from $2\div3$ (Jaffe 1983; Hernquist 1990), the most probable values 
for $\Big(\frac{\widetilde{M_D}}{M^*_{tot}}\Big)_t$ turn out to range from $0.57\div 0.80$, which
corresponds to $log\Big(\frac{M^*_{tot}}{M_B}\Big)_t=0.37\div0.69$, with 
$log\Big(\frac{M^*_{tot}}{M_B}\Big)_t=0.50$ at $b=2.5$.
The agreement with the J\o rgensen's histogram in Fig.5 
(J\o rgensen 1999), related to early type galaxies  in the central part
of the Coma cluster when
the same IMF is assumed, appears to be very good.

But one of the most important issues of the present theory appears to be related to the
physical reason which does cause the {\it tilt}.

In order to have the {\it tilt} we need to have the maximum of Clausius' virial energy.
This, in turn, requires to have the equipartition between the self- and the tidal- energy
of the stellar component. By considering the derivative of $\widetilde {V_B}$ of
eq.(\ref{virn}) in respect to $x$ and according to eq.(\ref{Mtilde}) and the constraints
(\ref{constr}), we conclude that
the dark matter mass has to increase steeper than
$(a_B/a_D)$. This means, in turn, that $\rho_D$ has to
decrease less than $1/r^2$ at the border of the bright mass, 
in order that {\it tidal energy} may overcome
the self-energy from this border forwards. But if we enter deeper into the fine
play of the exponents we are able to deduce a stricter constraint.

Going back to the dependence of $I_e$ on the mass ratio $m$ and $M_B$,
the message of the (\ref{ie})
is: $I_e$ depends on the cosmological history of the
galaxies, on dark matter distribution and on the baryonic fraction
which is inside.
But the ratio $L/M_B$, that is the {\it tilt}, is totaly independent of the cosmic
perturbation spectrum and of the mass ratio $m$. It turns
out to depend only on the dark matter density profile.
Indeed,
if we look at $$L\sim I_er_e^2\Longrightarrow\sim m^{i+2r} M_B^{I+2R}$$

where the exponents satisfy the following relationships:

$$i+2r=0$$
$$I+2R=2/(3-d)$$

Therefore, it is clear how the ratio $L/M_B$ loses its direct connection with the
cosmology given by $\gamma'$.
Moreover
in order to have the {\it positive (=observed) tilt}, we need:

$$2/(3-d)<1~\Longrightarrow 0<d<1$$

This condition is stricter than the necessary condition for the maximum: $d<2$

For $d=1\Longrightarrow$ the {\it tilt} disappears.

For $2>d>1\Longrightarrow$ the {\it tilt} appears but it is negative ( the opposite
of that observed).

Therefore, the slope of FP tells us a constraint on the density distribution of dark matter halo.
To have a positive {\it tilt} we need
the $DM$ mass has to increase steeper than
$(a_B/a_D)^2$ at the border of the bright mass. This means, in turn, that $\rho_D$ has to
decrease less than $1/r$. If the contrary occurs ($\Longrightarrow \rho_D$ 
decreases faster than $ 1/r$ but less than $1/r^2$) the {\it tilt} changes its sign.
That immediately underlines a problem with a NFW density profile concerning the
inner part of the
halo. 
The debate is still open. From the theoretical side what appears relevant is the
conclusion of a recent paper by M\H{u}cket \& Hoeft (2003) in which the constraint
on the exponent in the central dark halo region
has to be: $0\le d\le 0.5$, obtained by using Jeans' equations. From the observation
side we underline a very strict limit for the exponent $\gamma$ ($\gamma\le 0.8$),
completely in disagreement with the majority of simulations ($\gamma\geq 1$),  
determined by fitting the rotation curves obtained with high resolution tecnique, for a
sample of spiral galaxies in which
dark halos density profiles of {\it Zhao} kind have been used (Garrido, 2003). 

\subsection{On the limits of the model}

The problem of generalizing the results obtained in the case of a two-component
model with two power-law density profiles and with two homogeneous cores 
is still open.Some considerations may help in this future work.

A) The existence of Newton's first theorem
(see, Fig.1), which allows us to ignore
the mass distribution of the dark halo outside the $B$ volume as regards
the dynamical effects on the stellar subsystem
and then for the trend of Clausius' virial trace tensor $V_B$, which is the key of 
the whole theory.

B) The conclusion of the previous subsection that is the presence of the {\it tilt}
and its sign depends
on the gradient of dark matter distribution at the border of bright mass confinement.

Both together seem to converge on the main role which the central region of the dark
halos plays in the present theory. That may occur in a satisfactory way only if the scale radius of the dark halo,
$r_{oD}$
is not too small in respect to the dimension which contains the most of the stellar mass.
In other words, we will expect that the dark halo concentrations have to be not too high.

We show in Fig.3 and Fig.4 (Marmo, 2003) the same as in Fig.2 in the case of 
more
general models which belong to 
{\it Zhao} models. That in order to prove as
the appearence of Clausius' minimum, for suitable values
of the concentrations, seems a common feature
of the these most general and most realistic triaxial models 
where the density profiles are given by eq.(\ref{nfw})
for both the system components (e.g., NFW+Her, Fig.3; {\it cored} NFW+
Her, Fig.4).	   
But an other problem arises, with this kind of models. That is to recover,  
analytically, the interplay of the exponents which appear in  FP-quantities.

\begin{figure}[!h]
\begin{center}
\includegraphics[height=15cm,width=12cm]{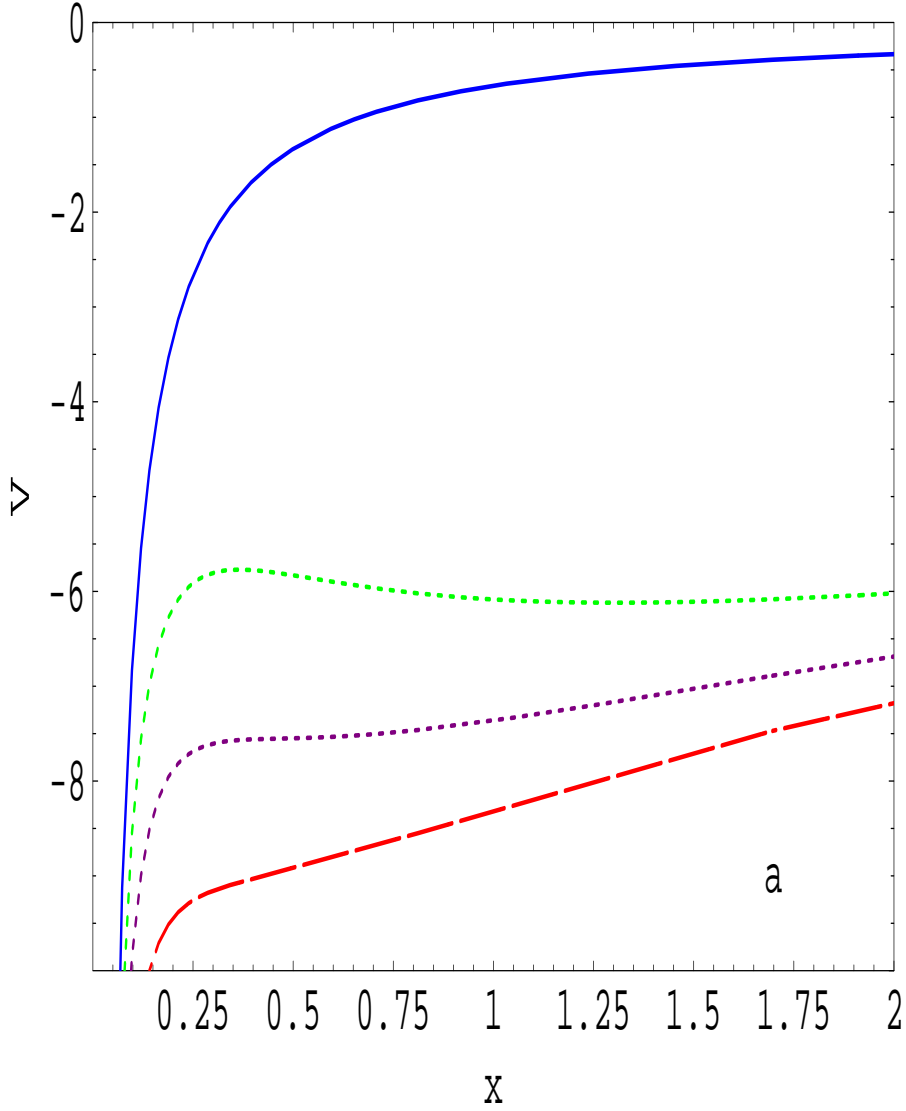}
\caption{The trends of the Clausius' virial energies in the normalized form, vs. $x$
defined as in Fig.2, (dashed curves), 
in the case of  two-component systems built up of a inner stellar component
with a Hernquist (1990) density profile and an outer dark halo with $NFW$ profile 
(eq.(\ref{nfw})).
The concentrations are: $C_B=4$ and, from top to bottom: $C_D=4,~7,~10$. The mass ratio is $m=12$.
The curve corresponding to:$C_B=C_D=4$, exibits surely a minimum for the absolute value of
Clausius' 
virial energy. 
For comparison, the self potential energy of the same $B$ component when single, is 
shown with a
continuous track (Marmo, 2003).}
\label{potenz}
\end{center}
\end{figure}

\begin{figure}[!h]
\begin{center}
\includegraphics[height=11cm,width=10cm]{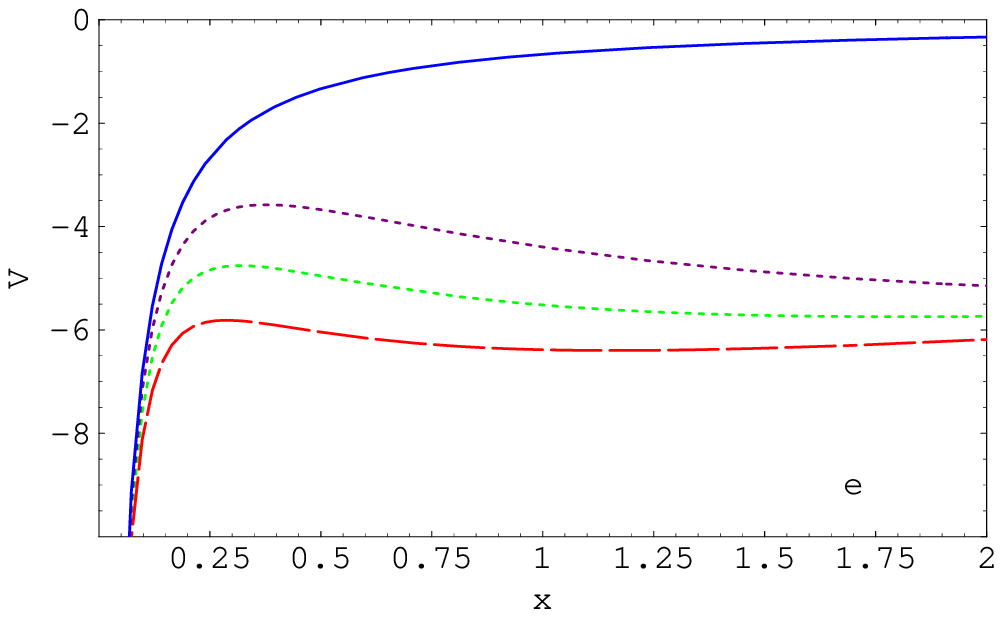}
\caption{The trends of the Clausius' virial energies in the normalized form, vs. $x$
defined as in Fig.2, (dashed curves), 
in the case of  two-component systems built up of a inner stellar component
with a Hernquist (1990) density profile ($\alpha=1;\beta=4;\gamma=1;\delta=3$) and an outer dark 
halo with a {\it cored} $NFW$ profile (eq.(\ref{nfw});$\alpha=1;\beta=3;\gamma=0;\delta=3$).
The concentrations are: $C_B=4$ and, from top to bottom: $C_D=4,~7,~10$. The mass ratio is $m=12$.
For comparison, the self potential energy of the same $B$ component when single, is 
shown with a
continuous track (Marmo, 2003).}
\label{potenz}
\end{center}
\end{figure}

\subsection{Thermodynamic arguments}
We will now consider the $B$ component thermodynamics inside a 
two-component system, to look for what characterizes
its special virial configuration from the thermodynamical point of view.
We begin to analyze the double system when it arrives at virialized
stages  
after a phase   
of violent relaxation ( Binney $\&$
Tremaine, 1987).
From this time  onwards, it may be assumed that the $B$ component
begins its virial evolution which consists of a sequence of slow contractions
witht enough time to rearrange the virial equilibrium after any step of
the sequence. 
In this way 
the thermodynamic process of contraction may be divided into
a sequence of transformations which are irreversible but 
occur between quasi-equilibrium stages with the typical character of
{\it external thermal irreversibility} (Zemansky, 1968).
Therefore, it is possible to assign a mean temperature $\overline T_S$
to the whole component during this quasi-static sequence of its dynamic evolution 
in the following way. 

\begin{figure}[!h]
\begin{center}
\includegraphics[height=15cm,width=12cm,angle=-90]{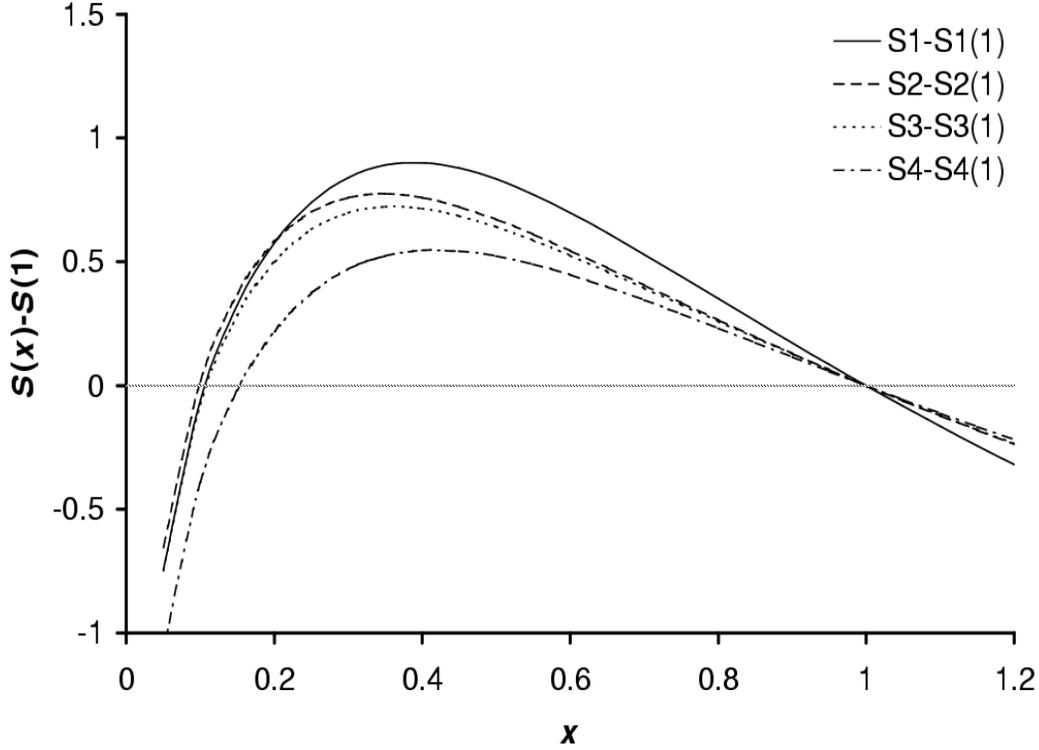}
\caption{Trends of the entropy function $\widetilde{F}(x)$ (eq.(\ref{fentro})) for the 
$B$-system, in arbitrary units (see text), as 
function of $x$ (see, Fig.2). The cases are of Tab.1. The maxima occur
at the {\it tidal radii} (Raffaele, 2003).}
\label{potenz}
\end{center}
\end{figure}

By assuming that the $B$ stellar system has
an isotropic velocity distribution, we may define its mean
temperature as:
\begin{equation}
\label{temp}
\overline {T}_S=\frac{m_*<\sigma^2>}{k}
\end{equation}
where $m_*$ 
is the mean mass of the stars and 
$k$ is the Boltzmann constant
(see, e.g., Lima Neto et
al. 1999; the a-$Ansatz$, in Bertin \& Trenti, 2003).

Moreover, if we limit ourselves to the macroscopic pressure supported elliptical systems, which are
relevant in order to define the FP we are dealing with, it follows that $\overline {T}_S$ is related to the
dominant peculiar kinetic energy, $\frac12M_B<\sigma^2>\simeq T_B$, in the following way:
\begin{equation}
\label{Temp2}
\overline {T}_S\simeq \frac{2T_B}{N~k}
\end{equation}
$N$ being the star number of the $B$ component.
$T_B$ is, in turn, connected with the central projected velocity
dispersion $\sigma_o$ by the usual factor $k_v$ which links
the kinematic galactic structure with $\sigma_o$ as follows (e.g., LS1):
\begin{equation}
\label{kv}
\sigma_o^2=k_v<\sigma^2>
\end{equation}
In the $I^o$ Thermodynamic Principle equation for the $B$ system what has to appear 
is the work done against its pressure  by both the forces on the 
system: the self 
gravity and the gravity which the dark matter distribution exerts on it. The corresponding 
potential energy
variation, for a small contraction, will be $\Delta V_B$ in such a way
as to have:
\begin{equation}
\label{ET}
\Delta E_T = \Delta Q-\Delta V_B
\end{equation}

where $E_T$ is the internal energy of $B$ and $\Delta Q$ is the small amount
of heat
the structure is able to exchange with the surrounding medium, 
in which the two-component
system lies,e.g., by {\it cooling} processes.
Indeed,
we have to take into account that the quasi-static evolution
of the inner component ( the $D$ one is considered non-dissipative) needs to involve some dissipative processes,
as Prigogine (1988) has well pointed out, on the general grounds. In a gas 
dominated structure,
dissipation may easily occur, e.g., by gas clouds collisions, but also in a 
collisionless
stellar component such as an elliptical.
Indeed, several evolutionary processes 
may potentially occur  
(see, e.g., Bertin \& Trenti, 2003), some of them with a dissipative character. 

If enough time is available in order to restore virial equilibrium, 
the variation of the virial 
quantities 
of the same component during this
quasi-static transition, turns out to be: 

\begin{equation}
\label{DVir}
\Delta T_B = -\Delta V_B/2
\end{equation}

Combining the two equations (\ref{ET}, \ref{DVir}) and considering that the 
internal energy
may now be identified with the total macroscopic kinetic energy of the stars, $T_B$,
according to eq.(\ref{Temp2}), we obtain
the two equations: 

\begin{eqnarray}
\label{varvir1}
\Delta Q=\Delta V_B/2\\
\label{varvir2}
\Delta \overline{T}_S\sim -\Delta V_B/2
\end{eqnarray}

The two requests that the system settles in virial 
equilibrium and
obeys the $I^o$ Principle yields the equipartition of the virial energy
variation which now becomes the variation of the Clausius' virial 
energy: 
that is one half of it has to be exchanged with {\it rest of the univese},
the other half has to contribute to change the 
temperature of the system. 
The result is formally the classical one found by Chandrasekhar (1939) and Scwarzschild (1958) 
for a single
gaseous stellar component.
The big difference, for the $B$ component 
inside the $D$ one,
is that the potential energy $V_B$ is now a non-monotonic function of $a_B$ (Fig.2).
 
 The variation of the entropy of the $B$ 
system, ${\bf S}_B$ , during the transformation between two virial states due to a 
small contraction $\Delta a_B$, which has the typical character of {\it external
thermal irreversibility}, is evaluated as:
\begin{equation}
\label{DS}
\Delta {\bf S}_B=\frac{1}{\overline{T}_S} \Delta Q =2Nk\frac{\Delta (V_B/2)}{-V_B/2}
\end{equation}

 If $\Delta V_B$ is negative due to the energy lost by radiation, the consequence 
 of the radiation flow is
 an entropy amount $\Delta {\bf S}_r$ of the thermal radiation bath  
 in the surroundings, which has the mean 
temperature
$T_r$, equal to:
\begin{equation}
\Delta {\bf S}_r=\frac12\frac{( -\Delta V_B)}{T_r}
\end{equation}
$\Delta {\bf S}_r$ is
	    of opposite sign
	    of $\Delta {\bf S}_B$ and greater than it, in absolute value, 
	    in a way that, according to the $II^o$ Thermodynamics Principle,
	    the {\it whole universe} increases its entropy $\bf {S}_u$ of:
	    \begin{equation}
	    \label{IIPrin}
	    \Delta {\bf S}_u=\frac12 (-\Delta V_B)(\frac{1}{T_r}-\frac{1}{\overline{T}_S}) 
	    \end{equation}
	   That holds because the surrounding thermal bath, in which the radiation
	   flow
	   will at the end thermalize, has a mean temperature lower than that of 
	   the stellar structure.

Due to the non-monotonic character of the $V_B$ trend, these very important 
consequences follow: 

A) as soon as the Clausius' virial energy is stationary
${\bf S}_B$ must also be so. That occurs at $a_t$ because the corresponding configuration
satisfies the d'Alembert Principle of {\it virtual works} (see, eq.(\ref{lavts})).
Therefore, one expects to obtain a maximum or a minimum
for the entropy of the $B$ component when the stellar system is on its
{\it tidal radius} configuration. It should be underlined that this result is 
independent of
the special class of two-component models we are dealing with 
but it depends only on the meaning of Clausius' virial energy and on
presence or not of the Clausius' virial maximum
during the quasi-static contraction sequence the chosen models are able 
to describe the $B$ subsystem evolution.
 Indeed, the eq.(\ref{DS}) derives from the physical 
meaning of Clausius' energy and from the first Thermodynamical Principle
together with the virial constraint.

B) An other consequence of eq.(\ref{DS}) is: all the configurations of the 
stellar system which correspond to a dimension greater than the {\it tidal
radius} $a_t$ are forbidden. Indeeed let us consider any configuration on 
the right side of the $x_t$ in Fig. 2. Starting from this one we take
into account the thermodynamical transformation which follows to a
small contraction $\Delta a_B<0$. According to the eqs.(\ref{Temp2}, \ref{varvir1},
\ref{varvir2}, \ref{DS}), the consequences are:

\begin{equation}
\label{forbid1}
\Delta a_B<0;~\Delta\overline{T}_S<0;~\Delta Q>0;~\Delta {\bf S}_B>0
\end{equation}
The stellar structure would have to increase its entropy by taking energy from the
radiation bath in which it is embedded  and which has a lower temperature than 
$\overline{T}_S$. It is manifest that this termodynamical process contradicts the
second Principle.
Starting from the same configuration, we now consider a small expansion.
We would obtain:
\begin{equation}
\label{forbid2}
\Delta a_B>0;~\Delta\overline{T}_S>0;~\Delta Q<0;~\Delta {\bf S}_B<0
\end{equation}
From the thermodynamical second Principle that would be possible but the contradiction
arises from the conservation of the total energy which is
equal to the half ot the total potential energy, $\frac12E_{pt}$, of the 
two-component system (see, MS3, Fig.5). Indeed a small expansion
would request an increase of the total mechanical energy of the whole system.

The main result is that all the virial configurations greater 
than $a_t$ are forbidden.

C)~On the contrary, all the configurations characterized by a semimajor axis
$a_B<~a_t$ correspond to obtain, for a small contraction, the transformations:
\begin{equation}
\label{allow1}
\Delta a_B<0;~\Delta\overline{T}_S>0;~\Delta Q<0;~\Delta {\bf S}_B<0
\end{equation}
which are allowed by the second Principle.
A small expansion would on the contrary yield to:
\begin{equation}
\label{forbid3}
\Delta a_B>0;~\Delta\overline{T}_S<0;~\Delta Q>0;~\Delta {\bf S}_B>0
\end{equation}
which are forbidden both from the energy conservation Principle and
from the second Thermodynamical Principle.
The last pair of relations are those we may obtain by using 
the eqs.(\ref{varvir1}, \ref{varvir2}, \ref{DS}) 
in  the case of one single $B$ component. That is clearly 
because: for $a_B<a_t$ the regime is characterized by the overcoming
of self-gravity on the tidal-gravity.

The main conclusion is: {\it the scale length induced on the stellar
component by the dark matter halo works as a real border of this
subsystem in the same way as the Hoerner's (1958) tidal radius, induced
by the galaxy, acts as a
confinement for the stellar of a globular cluster (see, Appendix A)}.
Therefore, it appears rasonable that ellipticals and GCs belong to the same
FP and that King's (1966) models, which need tidal cut-off,
could be shared by both kinds of objects (Djorgovski, 1995; Burstein et al.
1997; Secco, 2003). 

Moreover, the possible way in which the special 
{\it tidal radius} configuration may be reached during the violent
relaxation phase of the system, 
gains a deep meaning. Indeed, this configuration
is the widest one, the baryonic matter may have inside the dark 
potential well,
and corresponds to the minimum of the macroscopic random velocity pressure
the stellar system has to gain in order to virialize itself.

\section{Entropy trend and thermodynamic information}
By integration of eq.(\ref{DS}) (regarding the small variations
as infinitesimals) we may obtain the trend of
the entropy, normalized to the factor $2Nk$, $\widetilde {{\bf S}}(x)$, 
along the quasi-static virial 
sequence of the $B$ component, as follows
(see also, MS3):
\begin{equation}
\label{fentro}
\widetilde{{\bf S}}(x)-\widetilde {{\bf S}}(1)=ln\frac{V_B(1)}{V_B(x)}=\widetilde{F}(x);~~~~ x=a_B/a_D
\end{equation}
where $x=1$ is obtained when
both the two subsystems coincide.
The absolute value of $V_B(x)$ exibits a minimum at $x=x_t$ along
the virial sequence, therefore the function $\widetilde{F}(x)$ has its
maximum at the special configuration of $B$. The  corresponding
entropy variation, in physical units, along the sequence is simply $2Nk\widetilde{F}(x)$.

In the case of similar heterogeneous models considered, 
for the four cases of Tab.1, we obtain the 
trends shown in Fig.5.

\begin{figure}[!h]
\begin{center}
\includegraphics[height=15cm,width=12cm, angle=-90]{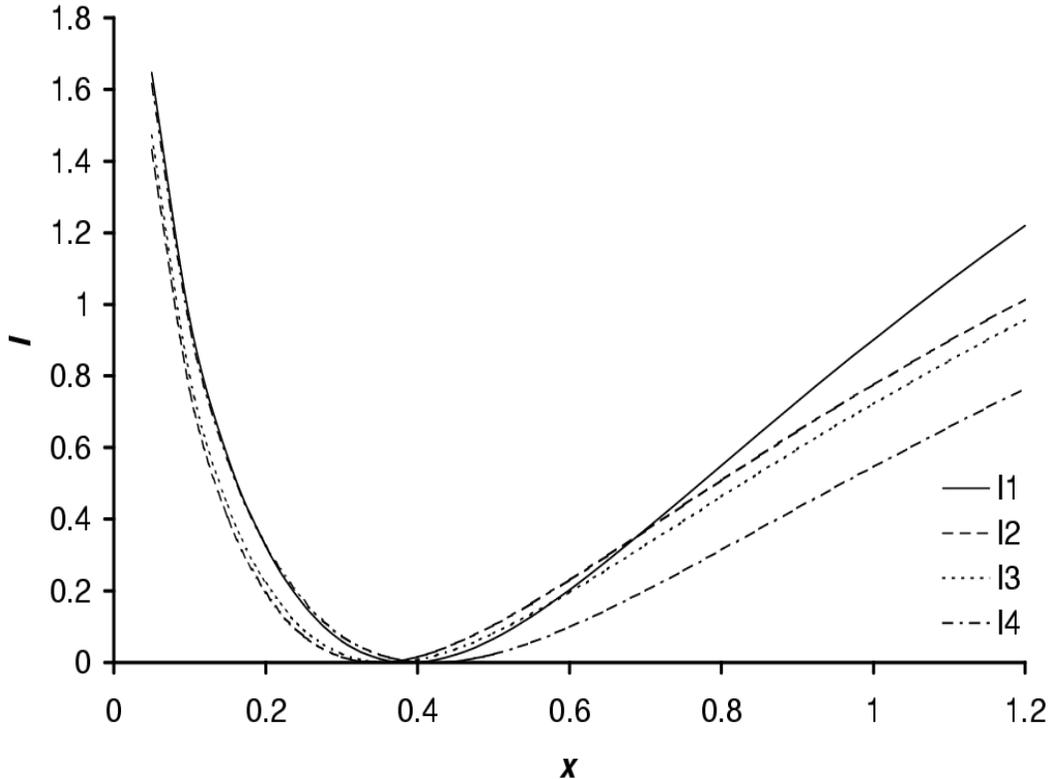}
\caption{Thermodynamical Information trends for the $B$-system, in arbitrary 
units (see text), as a 
function of $x$ (see Fig. 2) for the cases of Tab.1. The maximum  normalized value $\widetilde{{\bf S}}_{max}$ is
here assumed to be $\widetilde{{\bf S}}_{max}=\widetilde{{\bf S}}(x_t)$ instead of the real maximum
entropy value 
corresponding to the cosmic background radiation (see, eq.(\ref{infarm})) (Raffaele, 2003).}

\label{marmo3}
\end{center}
\end{figure}

The last thermodynamical quantity to be considered will be 
the thermodynamic information,
which, according to Layzer (1976), is:
\begin{equation}
\label{inf}
I={\bf S}_{max}-{\bf S}
\end{equation}
where ${\bf S}_{max}$ means the maximum value the entropy of the system may have as soon as the constraints
on it, which fix the actual value of its entropy to ${\bf S}$, are relaxed. 
Therefore, in order to 
increase its 
information a system has to decrease its entropy more and more in respect 
to that of the universe
(the maximum available and given essentially by the entropy of $CBR$; see, e.g.: Coles \& Lucchin, 1995; Secco 
1999). 

For a luminous 
component such as $B$, which is embedded in an other $D$, a contraction 
decreases the entropy 
function $\widetilde{F}(x)$  
only if its
dimension is smaller or equal to $a_t$.
If we now transform the information given by the eq.(\ref{inf}),
into the same units of $\widetilde{F}(x)$, we
obtain:
\begin{equation}
\label{infar}
\widetilde{I}_{min}=\widetilde{{\bf S}}_{max}-\widetilde{{\bf S}}(1)-\widetilde{F}(x_t)
\end{equation}

This means that from {\it tidal radius} downwards the 
component $B$ may gain information by contraction.
We plot in Fig.6 the trends of the information $I$
in the same units of $\widetilde{F}(x)$ by assuming $\widetilde{{\bf S}}_{max}
=\widetilde{{\bf S}}(x_t)$
instead of the real maximum
entropy value 
corresponding to the $CBR$. The eq.({\ref{inf}) becomes:

\begin{equation}
\label{infarm}
\widetilde{I}=\widetilde{{\bf S}}(x_t)-\widetilde{{\bf S}}(1)-\widetilde{F}(x)
\end{equation}
The {\it tidal radius} configuration appears again as the best
candidate for the beginning of the virial stage because it corresponds to
the minimum of thermodynamical information along the whole virial sequence. 
The stellar subsystem
has the real possibility to evolve onwards, with the 
dissipative processes indicated,
e.g., by Bertin \& Trenti (2003), by becoming more structured 
than it does at the end of the relaxation
phase, that is at the special configuration given by $x_t$. 
The same arguments may be transferred to the cut-off spirals 
which, in this frame,
had to begin a significant structure evolution from this stage forwards.
Even if this extension has to be made wider and deeper in the future, 
it may 
become the ground
for the interpretation of the cut-off radius  observed in many edge-on spiral galaxies
(see, e.g., van der Kruit, 1979;
Pohlen et al. 2000a, 2000b and Kregel et al., 2002). A preliminary
analysis has been done in Guarise et al. (2001) and in  Secco \& Guarise (2001). 

\section{Concluding remarks on the special configuration}
The consequence of the existence of a special
configuration characterized by the {\it tidal radius}
have been analyzed.
We stress particularly that:

a) at this special configuration the two-component models considered, built up of two homothetic similar strata spheroids  
with two power-laws
and two different homogeneous cores, yield
outputs which are relevant for some observable
scaling relations of pressure supported ellipticals (subsect. 4.3 and 4.4).
Moreover they could explain the physical reason for the existence of a 
Fundamental Plane for two-component virialized systems
(e.g.,Dressler et al.1987; Djorgovski \&
Devis, 1987; Bender et al. 1992; Djorgovski \& Santiago, 1993;
Burstein et al. 1997; Bertin, Ciotti \& Del Principe, 2002;
Borriello, Salucci \& Danese, 2003, and references therein).

b) The models
belong, once their profiles
are smoothed,
to the class with profiles of general pseudo-isothermal kind,
in turn, a sub-case of the more general class with  profiles introduced by Zhao (1996). 
Indeed the density distributions of a realistic two-component
model for an elliptical belong to this last class: both the dark matter halo profile, 
as proposed by 
Navarro, Frenk \& White  
(Navarro et al. 1996, Navarro et al. 1997) (see section 2) and the Hernquist (1990)
profile for the stellar component. 

Even if the analytical properties we consider here are
those we deduce by using the non-smoothed power-law profiles with two homogeneous cores,
they may be transfered, to a good extent,
to
the general pseudo-isothermal case. The other problem, to recover analytical, similar 
results also 
in the 
more realistic model with, e.g., NFW+Her profiles, is still open and has been 
adressed in the subsect. 4.5.

c) The  
thermodynamical relevance of this induced scale length is high in 
connection with the
Fundamental Plane of pressure supported ellipticals. Indeed, from the point
of view of the
two-component galaxy thermodynamics, deduced by the $I^o$ Thermodynamic Principle
under the virial equilibrium constraint,
this dimension works like a wall of a vessel, because a 
 larger configuration turns out to be forbidden
by the $II^o$ Principle of the Thermodynamics.
This cut-off on the luminous component space provides the gravitational field, which 
is intrinsically without
any  scale length, of a specific border, as it appears for other stellar systems such as 
the globular
clusters. As we have already shown (LS1, Secco, 2003; 
MS3) 
this truncation, which King (1966) has also introduced {\it ad hoc} in his 
primordial
models for ellipticals,
seems to be the common feature of all the astrophysical structures, from the 
globular clusters to the galaxy clusters, belonging to the {\it cosmic
metaplane}, (Djorgovski, 1995; Burstein et al. 1997). This is 
in the former induced by 
the Galaxy potential well, in the others by the dark matter halos. 
Its existence
might be able, in principle, to explain some of the 
fundamental scaling relations of these structures as we already have proved for the 
ellipticals (LS1) and, tentatively, for the GCs (Secco, 2003). 
How this special configuration may be reached, during the stellar system evolution, 
is strictly connected with
the problem of the end state of the collisionless stellar system after
a violent relaxation phase and then to the problem of the constraints
under which this phase occurs (Merritt, 1999). What we have already proposed in LS1,
is that the end state has to 
correspond to the minimum of the macroscopic pressure that the stellar system needs
in order to virialize. This is given by the maximum of Clausius' virial energy
configuration which, we will prove, also has the property of maximizing the entropy of the stellar
component located in a two-component virialized system. It 
corresponds to the wider configuration which does not break the second Thermodynamic
Principle together with energy conservation. This enforces the idea
of looking at this special configuration as the best candidate for the end of a
violent relaxation process of a stellar system when it occurs inside a dark massive halo
component. Indeed, the presence of this widest allowed configuration for the baryonic 
matter, which has the least requests for sustaining the structure
in virial equilibrium, would justify the fact that the ellipticals are not completely relaxed systems
 in respect to the collisionless dark halo.
 As White \& Narayan (1987) have pointed out, by studying single power-law
 stellar structures, the ellipticals seem indeed to have stopped their violent
 relaxation process before its end, unlike the collisionless dark matter
 structures. Really, the $NFW$ profile is given by eq.(\ref{nfw}) with the 
 exponent $\beta=3$ instead of the Hernquist profile, which is in agrement with
the de Vaucouleurs light profile, and requires $\beta=4$.

Moreover, it should be underlined
that, due to the physical meaning of Clausius' virial energy, the
maximization of entropy inside the virial evolution sequence is guaranteed as soon as the model exibits the minimum
of the absolute value of Clausius' virial, apart from the specific model we are
dealing with (see, subsect. 4.6).

\section{Conclusions}
The Mechanics and Thermodynamics of a stellar virialized system, embedded in a dark matter halo,
have been considered. The models (the same used in LS1)
are not derived by looking for the $DF$ which maximizes the standard Boltzmann-
Gibbs entropy. They consist of two heterogeneous spheroids with two power-law 
profiles and with two homogeneous cores. The concentrations are 
both equal $10$. Under some restrictions, they may be considered as belonging to a 
sub-class of the general class of models endowed with {\it Zhao density
profiles}. 
The most relevant aspects are:

$$(\alpha)$$- Under the usual constraints on the exponents of the density profiles
and the assumption of a frozen dark matter halo,
a configuration exists, characterized by having 
a dimension called {\it tidal radius} ($a_t$), which enjoys special properties
both from the mechanic and the thermodynamic point of view.

The mechanic properties at $a_t$-configuration are:

i)-the absolute value
of the Clausius' virial energy reaches its minimum.

ii)-The d'Alembert Principle of {\it virtual works} is satisfied, therefore
it is a configuration of equilibrium. Nevertheless it is not of {\it stable}
equilibrium because it does not minimize the total potential energy $(E_{pot})_B$
of the stellar subsystem (see, Fig.2 and MS3).

This special size translates to $FP$-like relation
by yielding outputs very relevant for some
observable scaling relations connected with
pressure supported ellipticals.

To fit the observed FP would require the dark halo density distribution
to be close to a generalized isothermal profile with an exponent $d$ around the
value $0.5$ at the border of Baryonic mass confinement. A stringent upper limit 
appears to be $d=1$ for which the gradient of dark matter distribution
would produce a {\it tilt} exponent: $\alpha_t=0$. That means: the observed {\it tilt}
disappears.

The thermodynamical properties at $a_t$-configuration are the following:

A)-The last mechanical property has the consequence, from the thermodynamical point of view,
that the work done from the external forces, i.e. the self gravity and the
tidal gravity, against the macroscopic pressure of the $B$ stellar subsystem,
is zero at $a_t$ for any small contraction or expansion starting from it. 
This also means that the entropy of $B$ is 
stationary at $a_t$, if one considers the first 
Thermodynamics Principle under
the virial equilibrium constraint for the thermodynamical irreversible
transformations occurring between two consecutive quasi-equilibrium stages 
in which
the evolutive sequence may be divided.
At every quasi-equilibrium configuration it is possible to associate a mean temperature
and at two consecutive stages, an entropy variation.

B)-By integration of the entropy variation, it appears 
that entropy reaches, inside the evolutive virial sequence, its maximum at $a_t$. 

Therefore, the first conclusion is that:
as soon as a class of models is able to exibit a minimun of the virial energy
along an evolutive sequence of the two-component structure they describe,
the model of this class which corresponds to the Clausius' minimum (in absolute value)
is also the
model which maximizes the entropy inside the virial sequence. 
This is verifyed in spite of the class of models we are dealing with 
and it appears new in respect to the results, for single structures, which appear 
in the literature. 
Moreover, this also allows us to avoid a pragmatic approach in which the 
thermodynamic
request is separated from the mechanical one as in Lima Neto et al. (1999).
Indeed the configuration at $a_t$ enjoys both
the best properties, mechanic or thermodynamic.

$$(\beta)$$- The second relevant aspect is that the mechanic and thermodynamic 
properties together reveal the confinement capability of the
scale length induced on the stellar subsystem from the dark matter halo,
That turns out to have the same role of
the tidal radius induced by the Galaxy on a globular
cluster, as  Hoerner (1958) found, and  represents a generalization of his result
(see, Appendix A).
This seems to be the key for understanding the physical reason for the
FP structure and why objects with a completely different history of
formation and evolution share the same FP features.
Indeed, the second Principle of Thermodynamics and
the energy conservation transform this {\it tidal radius} in a true
cut-off, i.e. the stellar system, as a whole, cannot have a 
configuration wider than that which corresponds to the special one.
That may be regarded as connected with the request, coming from many open 
problems, as referred in the introduction,
of a tidal cut-off on the coordinate space for the $DF$. 

$$(\gamma)$$-The third relevant aspect is related to the possible way
in which this special configuration may be reached. Since the beginning 
(LS1) we noted that Clausius virial maximum corresponds to the minimum of the
macroscopic pressure a subsystem needs for virialize during
the relaxation phase and the consequent conversion of radial ordered 
velocity into a dispersion velocity field (see, e.g., Huss et al. 1999).
This now becomes more than an $Ansatz$. Indeed, it appears that
the widest allowed configuration for the stellar system corresponds to this
minimum. 
This may explain how and
 why this special configuration will be reached by solving also the problem
 adressed by White \& Narayan (1987).
A possible physical reason is
the existence of this $a_t$-configuration for the baryonic component, 
needed because it is
less massive
than the dark halo and therefore more affected by the other component 
tidal effect 
than 
the halo does.

On the other hands the problem of knowing exactly the end state of a collisionless
stellar system after a violent relaxation phase, would require the knowledge
of the initial conditions under which this mechanism occurs (Merritt, 1999).
But as Djorgovski (1992) has pointed out and we have
shown in LS1\footnote{The degeneracy disappears for the ratio of the 
dark over baryonic 
mass , $m$, (see, LS1).} a degeneracy exists on the FP towards the 
cosmological initial conditions, then the possible end state of a stellar system
has probably to be
only indirectly deduced from the features of FP. The capability of Clausius virial
maximum configuration to explain some of the main features of the FP, may
highlight what is, otherwise, covered by intrinsic FP degeneracy.

$$(\delta)$$-The fourth is: taking into account the thermodynamical information during the
evolution sequence considered here for the $B$ stellar system, it arises
that the special configuration at $a_t$ corresponds to the minimum of this information.
That means an elliptical will gain small information due to its small
dissipation capability and therefore will stay at a low level of
structuration. On the contrary, a gas dominated galaxy will become more
structured and differentiated by this configuration forwards due to its
dissipation engine given, e.g., by cloud collisions.

\section*{Acknowledgements}

 My special thanks to Prof. R. Caimmi
 for his insightful suggestions and mathematical help, to Dr. Chiara Marmo, Dr.Andrea Raffaele 
 and Dr.Tiziano 
 Valentinuzzi 
 for having helped me
 in many ways as well as for the
 useful discussions. I am also grateful to the unknown referee
 for his patient contribution to focus, to improve the presentation
 and the understanding of the paper.
 
\appendix
\section{Appendix}
 
\subsection*{On the Hoerner's {\it tidal radius}}

How is the relationship between the {\it tidal} radius $a_H$ induced by the Galaxy on the 
satellites,
as the Globular Clusters (hereafter $GCs$), which Hoerner (1958) discovered and the
special size which maximizes the Clausius' virial energy of the Baryonic
component? The roles of confinement are the same (see, e.g., $(\beta)$-statements of
Conclusions) and the mathematical structures are very similar. 
The Hoerner's result is indeed:
\begin{equation}
\label{aH}
a_H=\Big( \frac{1}{2}
\frac{M_c}{M_G}\Big)^{\frac{1}{3}}~a_G
\end{equation}
where, $M_c$, is the GC mass and $M_G$ is Galaxy mass contained in the $a_G$  
radius. The eq.(\ref{aH})
has to be compared with the general result obtained in the contrentic case
given by
eq.(\ref{at}). 
$a_H$ is obtained by balancing of all the accelerations on the Cluster point $P$
in Fig.1A, when the variation of centrifugal force, due to a rigid Galaxy rotation,  
at the point $P$ is taken as negligible in respect to that of
the Cluster center $O$ (see, e.g., Brosche et al.1999). 
It should be underlined that the eq.(\ref{aH}) is 
derived in the Hoerner's point mass model. On the contrary, the eq.(\ref{at}) is 
deduced by the
non monotonic character of the Clausius' energy due to a mass extension of 
a component submitted both to
its self gravity and embedded in the gravity of an other one. 

The balance to consider is not of forces but between self- and tidal- potential energy with the 
same sign but with different trend. The question is: {\it in the concentric case only 
comparison between energy may be possible (the forces are concordant), but in the 
off-center case both the definitions are possible?} That means:~{\it in this last case also 
the Clausius'virial
of the Cluster shows a non-monotonic character?}

By analogy with the concentric case, we already guessed (Secco, 2003)
that, for some mass distributions of Cluster and Galaxy,
the answers were positive.
Detailed computations of the self- and the tidal-energy tensors
have been already performed in a previous paper (Caimmi \& Secco, 2003)
where the general second
order theory in $a_c/R_o$ ($R_o$ is the mean orbit radius of the $CG$) for 
the two off-center component systems
is developed.  
For heuristic sake, we take now into account the simple case 
in which both the two off-center 
components, Cluster and Galaxy,
are considered as two homogeneous spherical mass distributions 
(Caimmi \& Secco, 2005).

Generally speaking, the global tidal-potential energy trace due to the  
dynamical effect of the Galaxy-matter distribution on the cluster, $V_{cG}$, may be split into:
\begin{equation}
\label{split}
V_{cG}=V'_{cG}+ V_{cG}^o
\end{equation}
where $V_{cG}^o$ is the potential energy of a mass point placed at the cluster
barycentre, with same mass as the cluster, due to the Galaxy mass fraction $M_G(R_o)$ which is located 
inside the mean cluster distance $R_o$, 
and $V'_{cG}$ represents an additional term which is due to the cluster mass distribution.
The same holds for the global kinetic energy,
$T_c$, which, according to Koenig's theorem, may be split into two terms:
\begin{equation}
\label{cinetica}
T_c=T'_c+T^o
\end{equation}
one term, $T'_c$, is 
the intrinsic kinetic energy
of the cluster, the other one, $T^o$, is the kinetic energy of the cluster
barycentre.
Because of the time mean motion of the cluster barycentre occurs in stationary virial 
equilibrium, the following relation holds:
\begin{equation}
2T^o=-V_{cG}^o
\end{equation}
as soon as the two quantities are averaged over the period of cluster orbit inside the 
gravitational Galaxy field. Therefore, the virial equilibrium of the cluster mass distribution gives:
\begin{equation}
\label{virint}
2T'_c=-V'_c=-\Omega_c-V'_{cG}
\end{equation}
where $V'_{c}$ is the Clausius' virial energy of the cluster due to its mass distribution and 
$\Omega_c$ is its  
self-potential energy. 

 From the second
order theory in $a_c/R_o$, the following result may be deduced:
\begin {equation}
\label{tidalint}
V'_{cG}=-\frac{1}{3}G\frac{M_cM_G(R_o)}{a_c}\Big(\frac{a_c}{R_o}\Big)^{3}
\end{equation}

Then the Clausius' virial related to the mass distribution 
of the globular cluster becomes:
\begin{equation}
\label{Vprimc}
V'_c=-\frac{3}{5}\frac{GM_c^2}{a_c}-\frac{1}{3}G\frac{M_cM_G(R_o)}{a_c}\Big(\frac{a_c}{R_o}\Big)^{3}
\end{equation}
The non-monotonic character of $V'_c$ is manifest and then it follows that
also in the case of two 
off-center component system, the Clausius' virial 
exibits a maximum given by:
\begin{equation}
\label{aGt}
a_{Gt}=\Big( \frac{9}{5}\Big) ^{1/3}\Big (
\frac{1}{2}\frac{M_c}{M_G}\Big)^{\frac{1}{3}}~a_G 
\end{equation}
which turns out to be only about a factor $1.2$ greater than the $a_H$ radius.
The corresponding thermodynamical arguments in the case of two off-center 
component system are still an open question. 

\begin{figure}[!h]
\begin{center}
\includegraphics[height=10cm,width=12cm]{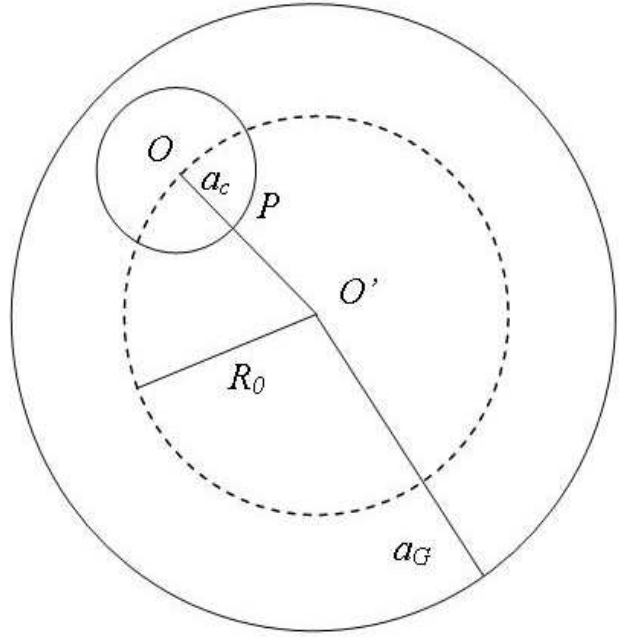}
\caption{Two off-center spherical component model. The satellite
(e.g., a Globular Cluster) is embedded in the gravity field of an other 
subsystem (e.g., the Galaxy). Its mass $M_c$ is distributed inside
$a_c$ and its barycenter $O$ moves in a mean circular 
orbit (dashed circle) of radius $R_o$  around the center $O'$ of the other
component with mass $M_G$ inside the radius $a_G$.  
$P$ is the point on which 
the accelerations hold the balance when $a_c$ becomes the {\it tidal radius}
of Hoerner.}

\label{potenz}
\end{center}
\end{figure}

 \section{Appendix }
 \subsection*{Glossary of Simbols (in order of comparison)}
\begin{itemize}

\item~~~~~~~~~~~~~~{\bf Section 1}

\item~~~{\bf S}~~~~~~~~~~~~~~~~~~~entropy
\item~~~$f(\vec{x},\vec{v})$~~~~~~~~~~~~~distribution function, $DF$

\item~~~~~~~~~~~~~~{\bf Section 2}

\item~~~ $\rho(r)$~~~~~~~~~~~~~~~density radial profile
\item~~~$\alpha, \beta, \gamma, \delta$~~~~~~~~~~~exponents of {\ Zhao density profile}
\item~~~$B$~~~~~~~~~~~~~~~~~~~Bright (stellar) or Baryonic component
\item~~~$DM$~~~~~~~~~~~~~~~~Dark Matter
\item~~~$D$~~~~~~~~~~~~~~~~~~~Dark Matter component
\item~~~$(T_B)_{ij}$~~~~~~~~~~~~~~kinetic-energy tensor of Baryonic component
\item~~~$(V_B)_{ij}$~~~~~~~~~~~~~~Clausius'virial tensor of Baryonic component
\item~~~$(T_D)_{ij}$~~~~~~~~~~~~~~kinetic-energy tensor of $DM$ component
\item~~~$(V_D)_{ij}$~~~~~~~~~~~~~~Clausius'virial tensor of $DM$ component
\item~~~$(\Omega_B)_{ij}$~~~~~~~~~~~~~~self potential-energy tensor of Baryonic component
\item~~~$(V_{BD})_{ij}$~~~~~~~~~~~~~tidal potential-energy tensor of Baryonic component 

~~~~~~~~~~~~~~~~~~~~~~~~~~~~~~~~~~~~~~~~~~~~due to the gravitational $DM$ force on it
\item~~~$(\Omega_D)_{ij}$~~~~~~~~~~~~~~self potential-energy tensor of $DM$ component
\item~~~$(V_{DB})_{ij}$~~~~~~~~~~~~~tidal potential-energy tensor of $DM$ component due 

~~~~~~~~~~~~~~~~~~~~~~~~~~to the gravitational force of Baryonic component on it
\item~~~$\Sigma^*$~~~~~~~~~~~~~~~~~~~~the surface which contains the $DM$ mass fraction which, 

~~~~~~~~~~~~~~~~~~~~~~according to Newton's $I^o$ theorem, exerts dynamical 

~~~~~~~~~~~~~~~~~~~~~~effect on the stellar component
\item~~~$\widetilde{M_D}$~~~~~~~~~~~~~~~~~approximate $DM$ mass fraction inside $\Sigma^*$

\item~~~$\Phi_B$~~~~~~~~~~~~~~~~~~gravitational potential due to the Baryonic 

~~~~~~~~~~~~~~~~~~~~~~~~~~~~~~~~~~~~~~~~~~~~~~~~~~~~~~~~~~mass distribution
\item~~~$\Phi_D$~~~~~~~~~~~~~~~~~~gravitational potential due to $DM$ mass distribution
\item~~~$\Phi_{ij}$~~~~~~~~~~~~~~~~~~gravitational tensor potential
\item~~~$G$~~~~~~~~~~~~~~~~~~~~Gravitational constant
\item~~~$T'_{ij}$~~~~~~~~~~~~~~~~~~~kinetic-energy tensor of ordered velocities
\item~~~$\Pi_{ij}$~~~~~~~~~~~~~~~~~~two times kinetic-energy tensor of random velocities
\item~~~$\nu (\vec{x})$~~~~~~~~~~~~~~~~spatial density in phase space
\item~~~$b$~~~~~~~~~~~~~~~~~~~~~exponent of power-law density profile in the cored 

~~~~~~~~~~~~~~~~~~~~~~~~~~~~~~~~~~~~~~~~~~~~~~~~~~~~~~~~~~~~~~Baryonic component 

\item~~~$d$~~~~~~~~~~~~~~~~~~~~~exponent of power-law density profile in the cored $DM$

~~~~~~~~~~~~~~~~~~~~~~~~~~~~~~~~~~~~~~~~~~~~~~~~~~~~~~~~~~~~~~~~~~~~~~~~~~~~~~component
\item~~~$a$~~~~~~~~~~~~~~~~~ ~~major semiaxis of the virialized subsystem considered
\item~~~$M$~~~~~~~~~~~~~~~~~~mass of the virialized subsystem considered
\item~~~$\rho_o$~~~~~~~~~~~~~~~~~~~characteristic density of the virialized subsystem 

~~~~~~~~~~~~~~~~~~~~~~~~~~~~~~~~~~~~~~~~~~~~~~~~~~~~~~~~~~~~~~~~~~~~~~~~~~~considered
\item~~~$r_o$~~~~~~~~~~~~~~~~~~scale radius of the subsystem considered
\item~~~$C$~~~~~~~~~~~~~~~~~~concentration of the subsystem considered
\item~~~$r_e$~~~~~~~~~~~~~~~~~~the effective radius

\item~~~$I_e$~~~~~~~~~~~~~~~~~~mean effective surface brightness within $r_e$
		
\item~~~$\sigma_o$~~~~~~~~~~~~~~~~~~the central projected velocity dispersion

\item~~~~~~~~~~~~~~{\bf Section 3}

\item~~~$(E_{pot})_B$~~~~~~~~~~~total potential energy of the stellar component
\item~~~$W_{BD}$~~~~~~~~~~~~~~interaction-energy between the Baryonic and the 

~~~~~~~~~~~~~~~~~~~~~~~~~~~~~~~~~~~~~~~~~~~~~~~~~~~~~~~~~~~~~~$DM$ components
\item~~~$(W_{BD})_{ij}$~~~~~~~~~interaction-energy tensor between the Baryonic and the 

~~~~~~~~~~~~~~~~~~~~~~~~~~~~~~~~~~~~~~~~~~~~~~~~~~~~~~~~~~~~~~$DM$ components
\item~~~$a_t$~~~~~~~~~~~~~~~~~~tidal radius of the Baryonic component
\item~~~$\nu_{\Omega B}$~~~~~~~~~~~~~~~coefficient of mass-distribution in the Baryonic

~~~~~~~~~~~~~~~~~~~~~~~~~~~~~~~~~~~~self potential-energy tensor
\item~~~$\nu_{\Omega D}$~~~~~~~~~~~~~~~coefficient of mass-distribution in the $DM$

~~~~~~~~~~~~~~~~~~~~~~~~~~~~~~~~~~~~~~~~self potential-energy tensor
\item~~~$\nu'_V$~~~~~~~~~~~~~~~~~coefficient which weights the tidal potential-energy

~~~~~~~~~~~~~~~~~~~~~~~~~~~~~~~~~~~~in respect to the self potential-energy in the 

~~~~~~~~~~~~~~~~~~~~~~~~~~~~~~~~~~~~Clausius' virial of the Baryonic component
\item~~~$m$~~~~~~~~~~~~~~~~~~mass ratio of $DM$ to Baryonic component
\item~~~$\widetilde{m}$~~~~~~~~~~~~~~~~~~ratio of $DM$ mass fraction $\widetilde{M_D}$, 
to total stellar mass
\item~~~$M_{tot}^*$~~~~~~~~~~~~~~~total dynamical mass inside the size  of Baryonic

~~~~~~~~~~~~~~~~~~~~~~~~~~~~~~~~~~~~~~~~~~~~~~~~~~~~~~~~~~~~~~~~~~~~~~~~component
\item~~~$F$~~~~~~~~~~~~~~~~~~form factor, the same for $B$ and $DM$ components
\item~~~$x$~~~~~~~~~~~~~~~~~~~size ratio of Baryonic to $DM$ components : $x=a_B/a_D$ 
\item~~~$\widetilde{V_B}$~~~~~~~~~~~~~~~~Clausius'virial energy normalized to $a_D/(GM^2_BF)$

\item~~~~~~~~~~~~~~{\bf Section 4}

\item~~~$L_s$~~~~~~~~~~~~~~~~work done by the self gravity forces
\item~~~$L_t$~~~~~~~~~~~~~~~~work done by the tidal gravity forces
\item~~~$F(\vec{x})$~~~~~~~~~~~~~{\it extrinsic attribute} in Chandrasekhar's analysis;

~~~~~~~~~~~~~~~~~~~~~~~~e.g., gravitational potential, its first derivative, etc.

~~~~~~~~~~~~~~~~~~~~~~~~which are not {\it intrinsic} to the fluid element but  

~~~~~~~~~~~~~~~~~~~~~~~~related to its spatial location
\item~~~$S_o$~~~~~~~~~~~~~~~~initial volume of the inner system
\item~~~$\Delta S_o$~~~~~~~~~~~~~~small variation of the initial volume of the inner system
\item~~~$\vec{\xi}(\vec{x},t)$~~~~~~~~~~~generic small displacement, at the time $t$, of the fluid 

~~~~~~~~~~~~~~~~~~~~~~~~~~~~~~element from the unperturbed position $\vec{x}_o$, in the 

~~~~~~~~~~~~~~~~~~~~~~~~~~~~~~~~~~~~~~~~~~~~~~~~~~~~~~~~~~~~~~~~~~~~~~~inner component

\item~~~$\Delta F$~~~~~~~~~~~~~~Lagrangian change of $F$ due to the displacement $\vec{\xi}$
\item~~~$M_o$~~~~~~~~~~~~~~~~initial volume of the outer system
\item~~~$\vec{\xi'}(\vec{x},t)$~~~~~~~~~~~generic small displacement, at the time $t$, of the fluid 

~~~~~~~~~~~~~~~~~~~~~~~~~~~~~~element from the unperturbed position $\vec{x}_o$, in the

~~~~~~~~~~~~~~~~~~~~~~~~~~~~~~~~~~~~~~~~~~~~~~~~~~~~~~~~~~~~~~~~~~~~~~outer system
\item~~~$FP$~~~~~~~~~~~~~~~~Fundamental Plane
\item~~~$A$~~~~~~~~~~~~~~~~~~~exponent of $\sigma_o$ in the equation of $FP$
\item~~~$B$~~~~~~~~~~~~~~~~~~~exponent of $I_e$ in the equation of $FP$
\item~~~$L$~~~~~~~~~~~~~~~~~~~Luminosity
\item~~~$c_1$~~~~~~~~~~~~~~~~~~$L/(I_er_e^2)$
\item~~~$<\sigma^2>$~~~~~~~~~~~~mean square velocity dispersion of the stellar system
\item~~~$k_v$~~~~~~~~~~~~~~~~~~$\sigma_o^2/<\sigma^2>$
\item~~~$<R>$~~~~~~~~~~~~the gravitational radius
\item~~~$k_R$~~~~~~~~~~~~~~~~~$r_e/<R>$
\item~~~$c_2$~~~~~~~~~~~~~~~~~~$(Gk_Rk_v)^{-1}$
\item~~~$T_{pec}$~~~~~~~~~~~~~~~~~peculiar kinetic energy dominant in the stellar component
\item~~~$T_{rot}$~~~~~~~~~~~~~~~~~~rotational kinetic energy negligible in the stellar component
\item~~~$\alpha_t$~~~~~~~~~~~~~~~~{\it tilt} exponent of the 
$FP$: $M_B/L\sim M_B^{\alpha_t}$
\item~~~$n_{rec}$~~~~~~~~~~~~~~~~~~effective spectral index of perturbations in $CDM$ scenario
\item~~~$\gamma'$~~~~~~~~~~~~~~~~~exponent in the scaling relation for virialized haloes:

~~~~~~~~~~~~~~~~~~~~~~~~~~~~~~~~~~~~~~~~~~~~~~$a_D\sim M_D^{1/\gamma'}; \gamma'=6/(5+n_{rec})$
\item~~~$\alpha_{rec}$~~~~~~~~~~~~~~local slope of the $CDM$ mass variance
\item~~~$r$~~~~~~~~~~~~~~~~~~exponent of $m$ in the scaling relation:~$ r_e\sim m^rM_B^R$;

~~~~~~~~~~~~~~~~~~~~~~~~~~~~~~~~~~~~~~~~~~~~~~~~~~~~$r=\frac{(3-d)-\gamma'}{\gamma'(3-d)}$

\item~~~$R$~~~~~~~~~~~~~~~~~~exponent of $M_B$ in the scaling relation:~$ r_e\sim m^rM_B^R$;

~~~~~~~~~~~~~~~~~~~~~~~~~~~~~~~~~~~~~~~~~~~~~~~~~~~~~~$R=1/\gamma'$

\item~~~$i$~~~~~~~~~~~~~~~~~~exponent of $m$ in the scaling relation:~$I_e\sim m^i M_B^I
$;

~~~~~~~~~~~~~~~~~~~~~~~~~~~~~~~~~~~~~~~~~~~~~~~~~~~~~~$i=2\frac{\gamma'-(3-d)}{\gamma'(3-d)}$
\item~~~$I$~~~~~~~~~~~~~~~~~~exponent of $M_B$ in the scaling relation:~$I_e\sim m^i M_B^I$;

~~~~~~~~~~~~~~~~~~~~~~~~~~~~~~~~~~~~~~~~~~~~~~~~~~~~~~$I=2\frac{\gamma'-(3-d)}{\gamma'(3-d)}$
\item~~~$s$~~~~~~~~~~~~~~~~~~exponent of $m$ in the scaling relation:~$ \sigma_o\sim m^sM^S_B$;

~~~~~~~~~~~~~~~~~~~~~~~~~~~~~~~~~~~~~~~~~~~~~~~~~~~~$s=-\frac{1}{2}\frac{(3-d)-\gamma'}{\gamma'(3-d)}$

\item~~~$S$~~~~~~~~~~~~~~~~~~exponent of $M_B$ in the scaling relation:~$ \sigma_o\sim m^sM^S_B$;

~~~~~~~~~~~~~~~~~~~~~~~~~~~~~~~~~~~~~~~~~~~~~~~~~~~~~~$S=\frac{1}{2}\frac{\gamma'-1}{\gamma'}$
\item~~~$FJ$~~~~~~~~~~~~~~~Faber-Jackson relation
\item~~~$(M_{tot}^*)_t$~~~~~~~~~~total dynamical mass inside the size corresponding to

~~~~~~~~~~~~~~~~~~~~~~~~~~~~~~~~~~~~~~~~~~~{\it tidal radius} of Baryonic component
\item~~~$m_*$~~~~~~~~~~~~~~~~mean star mass in the $B$ stellar system
\item~~~$k$~~~~~~~~~~~~~~~~~~the Boltzmann constant
\item~~~$\overline {T}_S$~~~~~~~~~~~~~~~~mean temperature of $B$ stellar system:~$
                                            \frac{m_*<\sigma^2>}{k}$
\item~~~$N$~~~~~~~~~~~~~~~~~star number of the $B$ stellar component
\item~~~$E_T$~~~~~~~~~~~~~~~~internal energy of the $B$ component
\item~~~$\Delta Q$~~~~~~~~~~~~~~~small amount of heat exchanged between the $B$ system and  

~~~~~~~~~~~~~~~~~~~~~~~~~~~~~~~~~~~~~~~~~~~~~~~~~~~~~~~~the surrounding medium
\item~~~$\Delta {\bf S}_B$~~~~~~~~~~~~~~~small entropy variation of $B$ system

~~~~~~~~~~~~~~~~~~~~during the small quasi-static variation of its size $\Delta a_B$

\item~~~$\Delta {\bf S}_r$~~~~~~~~~~~~~~~ corresponding small entropy variation 

~~~~~~~~~~~~~~~~~~~~of thermal radiation bath in the surroundings

\item~~~$T_r$~~~~~~~~~~~~~~~~~~~~~~~~~~mean temperature of the surrounding 
thermal bath

\item~~~$\Delta {\bf S}_u$~~~~~~~~~~~~~~~small entropy variation of the {\it whole universe}

\item~~~~~~~~~~~~~~{\bf Section 5}

\item~~~$\widetilde{{\bf S}}(x)$~~~~~~~~~~~~~~~entropy of the $B$ 
system, normalized to 

~~~~~~~~~~~~~~~~~~~~~the factor $2Nk$, at the size ratio $x$ of Baryonic 

~~~~~~~~~~~~~~~~~~~~~to $DM$ components

\item~~~$\widetilde{{\bf S}}(1)$~~~~~~~~~~~~~~~entropy of $B$ system, normalized to
   
~~~~~~~~~~~~~~~~~~~~~the factor $2Nk$, when the size of Baryonic component is 

~~~~~~~~~~~~~~~~~~~~~the same of the $DM$ one 

\item~~~$\widetilde{F}(x)$~~~~~~~~~~~~~~~~~variation of normalized entropy $=\widetilde{{\bf S}}(x)-
\widetilde{{\bf S}}(1)=ln\frac{V_B(1)}{V_B(x)}$

\item~~~$I={\bf S}_{max}-{\bf S}$~~~~thermodynamical information
\item~~~$\widetilde{I}$~~~~~~~~~~~~~~~~~~~~thermodynamical information normalized to
   
~~~~~~~~~~~~~~~~~~~~~the same arbitrary units of $\widetilde{{\bf S}}$ 

\item~~~~~~~~~~~~~~{\bf Appendix A}

\item~~~$a_H$~~~~~~~~~~~~~~~~~~Hoerner's {\it tidal radius}
\item~~~$GC$~~~~~~~~~~~~~~~~~Globular Cluster
\item~~~$a_c$~~~~~~~~~~~~~~~~~~~$GC$ radius
\item~~~$a_G$~~~~~~~~~~~~~~~~~~Galaxy radius
\item~~~$M_c$~~~~~~~~~~~~~~~~~~$GC$ mass
\item~~~$M_G$~~~~~~~~~~~~~~~~~Galaxy mass
\item~~~$V_{cG}$~~~~~~~~~~~~~~~~~global tidal-energy of the $GC$ due to the Galaxy
\item~~~$V'_{cG}$~~~~~~~~~~~~~~~~~tidal-energy of the $GC$ due to its mass distribution

~~~~~~~~~~~~~~~~~~~~~~~~~~~~~~inside the Galaxy gravity field 
\item~~~$V_{cG}^o$~~~~~~~~~~~~~~~~~potential energy due to the Galaxy gravity field of the

~~~~~~~~~~~~~~~~~~~~~~~~~~~~~~~~~~~~~~~~~~~~~~~~~~~~~~~~~~~~~~~~~cluster barycenter
\item~~~$T_c$~~~~~~~~~~~~~~~~~~~global kinetic energy of the $GC$
\item~~~$T'_c$~~~~~~~~~~~~~~~~~~~intrinsic kinetic energy of the $GC$
\item~~~$T^o$~~~~~~~~~~~~~~~~~~kinetic energy of the cluster barycenter
\item~~~$\Omega_c$~~~~~~~~~~~~~~~~~~self-potential energy of the $GC$
\item~~~$V'_{c}$~~~~~~~~~~~~~~~~~~Clausius' virial energy of the cluster mass distribution
\item~~~$R_o$~~~~~~~~~~~~~~~~~~mean orbit radius of the $GC$
\item~~~$M_G(R_o)$~~~~~~~~~~~Galaxy mass fraction inside the sphere of radius $R_o$
\item~~~$a_{Gt}$~~~~~~~~~~~~~~~~~{\it tidal radius} at the maximum of Clausius' virial

~~~~~~~~~~~~~~~~~~~~~~~~~~~~in the two off-center spherical and homogeneous 

~~~~~~~~~~~~~~~~~~~~~~~~~~~~component model

\end{itemize}


\end{document}